\documentclass[pre,
 reprint,
superscriptaddress,
 amsmath,amssymb,
 aps,
]{revtex4-2}
\usepackage{rotating}
\usepackage{multirow}
\usepackage{comment}
\usepackage{graphicx}
\usepackage{dcolumn}
\usepackage{bm}
\usepackage{ulem}
\usepackage{mathtools}
\usepackage{siunitx}
\usepackage{pifont}
\usepackage{hyperref}

\usepackage[dvipsnames]{xcolor}

\newcommand{\myeqref}[2]{(\hyperref[#1]{\ref*{#1}#2})}
\newcommand{\myref}[2]{\hyperref[#1]{\ref*{#1}(#2)}}
\newcommand{\myrefletter}[2]{\hyperref[#1]{(#2)}}
\newcommand{\myrefnb}[2]{\hyperref[#1]{\ref*{#1}#2}}
\newcommand{\subtag}[1]{\tag{\theparentequation #1}}

\newcommand*\de{\mathop{}\!\mathrm{d}}
\newcommand*{\Oh}{\mathcal{O}}
\newcommand*{\ie}{\emph{i.e.}}
\newcommand*{\eg}{\emph{e.g.}}

\newcommand\figbcircle[1][.55\baselineskip]{\mathbin{\vcenter{\hbox{\resizebox{!}{#1}{${\bullet}$}}}}}
\newcommand\figwcircle[1][.55\baselineskip]{\mathbin{\vcenter{\hbox{\resizebox{!}{#1}{${\circ}$}}}}}

\newcommand\figwsquare[1][.5\baselineskip]{\mathbin{\vcenter{\hbox{\resizebox{!}{#1}{${\square}$}}}}}

\newcommand\figwtriangle[1][.5\baselineskip]{\mathbin{\vcenter{\hbox{\resizebox{!}{#1}{${\triangle}$}}}}}
\newcommand\figtimes[1][.5\baselineskip]{\mathbin{\vcenter{\hbox{\resizebox{!}{#1}{${\bm{\times}}$}}}}}
\newcommand\figplus[1][.45\baselineskip]{\mathbin{\vcenter{\hbox{\resizebox{!}{#1}{$\bm{+}$}}}}}

\newcommand{\beq}{\begin{equation}}
\newcommand{\eeq}{\end{equation}}

\newcommand*{\ThetaT}{\Theta_+}
\newcommand*{\ThetaB}{\Theta_-}
\newcommand*{\ThetaTB}{\Theta_\pm}

\newcommand*{\LT}{L_+}
\newcommand*{\LB}{L_-}
\newcommand*{\LTB}{L_\pm}
\newcommand*{\LLat}{L_\mathrm{L}}

\newcommand*{\hT}{h_+}
\newcommand*{\hB}{h_-}
\newcommand*{\hTB}{h_\pm}
\newcommand*{\hSur}{h_\mathrm{S}}
\newcommand*{\hLat}{h_\mathrm{L}}

\newcommand*{\ET}{E_+}
\newcommand*{\EB}{E_-}
\newcommand*{\ETB}{E_\pm}
\newcommand*{\ESur}{E_\mathrm{S}}
\newcommand*{\ELat}{E_\mathrm{L}}

\newcommand*{\TTB}{T_\pm}

\newcommand*{\NTB}{N_\pm}

\newcommand*{\RE}{\mathcal{R}_E}
\newcommand*{\Rh}{\mathcal{R}_h}
\newcommand*{\RL}{\mathcal{R}_L}
\newcommand*{\RT}{\mathcal{R}_{\Theta}}

\newcommand*{\hatA}{\hat{A}}

\newcommand*{\hatw}{\hat{w}}
\newcommand*{\hatt}{\hat{t}}
\newcommand*{\hatmu}{\hat{\mu}}

\newcommand*{\A}{\mathcal{A}}
\newcommand*{\hatb}{\mathcal{B}}
\newcommand*{\hatp}{\mathcal{P}}

\newcommand*{\hatwT}{\hat{w}_+}
\newcommand*{\hatwB}{\hat{w}_-}
\newcommand*{\hatwTB}{\hat{w}_\pm}

\newcommand*{\hatpT}{\mathcal{P}_+}

\newcommand*{\hatpTB}{\mathcal{P}_\pm}

\newcommand*{\hatYT}{\mathcal{Y}_+}
\newcommand*{\hatYB}{\mathcal{Y}_-}
\newcommand*{\hatYTB}{\mathcal{Y}_\pm}

\newcommand*{\hatbT}{\mathcal{B}_+}
\newcommand*{\hatbB}{\mathcal{B}_-}
\newcommand*{\hatbTB}{\mathcal{B}_\pm}

\newcommand*{\AT}{\mathcal{A}_+}
\newcommand*{\AB}{\mathcal{A}_-}
\newcommand*{\ATB}{\mathcal{A}_\pm}

\newcommand*{\Keff}{K_\mathrm{2D}}

\newcommand*{\Yeff}{Y_\mathrm{eff}}

\newcommand*{\Beff}{B_\mathrm{eff}}
\newcommand*{\hatBeff}{\hat{B}_\mathrm{eff}}
\newcommand*{\kapeff}{\kappa_\mathrm{eff}}
\newcommand*{\hatkapeff}{\hat{\kappa}_\mathrm{eff}}

\newcommand*{\Thetam}{\Theta_\mathrm{max}}

\newcommand*{\ellbg}{\ell_\mathrm{bg}}
\newcommand*{\ellstrain}{\ell_\mathrm{bend}}
\newcommand*{\ellstretch}{\ell_\mathrm{stretch}}
\newcommand*{\ellthick}{\ell_\mathrm{thick}}
\newcommand*{\ellwidth}{\ell_\mathrm{width}}
\newcommand*{\elllength}{\ell_\mathrm{length}}

\begin{document}

\preprint{APS/123-QED}

\title{Mechanics of pressurized cellular sheets}

\author{Thomas G. J. Chandler}
 \email{tgchandler@wisc.edu}
\affiliation{Mathematical Institute, University of Oxford, Woodstock Rd, Oxford, OX2 6GG, UK}
 \affiliation{Department of Mathematics, University of Wisconsin--Madison, Madison, WI 53703, USA}
 
 \author{Jordan Ferria} 
\affiliation{LadHyX, CNRS, Ecole Polytechnique, Institut Polytechnique de Paris, 91128, Palaiseau Cedex, France}

\author{Oliver Shorthose}
\affiliation{Department of Engineering Science, University of Oxford, Parks Road, Oxford, OX1 3PJ, UK}

\author{Jean-Marc Allain}
\affiliation{LMS, CNRS, Ecole Polytechnique, Institut Polytechnique de Paris, 91128, Palaiseau Cedex, France}
\affiliation{Institut Nationale de Recherche en Informatique et en Automatique, 91128, Palaiseau, France}

\author{Perla Maiolino}
\affiliation{Department of Engineering Science, University of Oxford, Parks Road, Oxford, OX1 3PJ, UK}

\author{Arezki Boudaoud} 
\affiliation{LadHyX, CNRS, Ecole Polytechnique, Institut Polytechnique de Paris, 91128, Palaiseau Cedex, France}

\author{Dominic Vella}%
 \email{dominic.vella@maths.ox.ac.uk}
\affiliation{Mathematical Institute, University of Oxford, Woodstock Rd, Oxford, OX2 6GG, UK}

\date{\today}

\begin{abstract}
Everyday experience shows that cellular sheets are stiffened by the presence of a pressurized gas: from bicycle inner tubes to bubble wrap, the presence of an internal pressure increases the stiffness of otherwise floppy structures. The same is true of plants, with turgor pressure (due to the presence of water) taking the place of gas pressure; indeed, in the absence of water, many plants wilt. However, the mechanical basis of this stiffening is somewhat opaque: simple attempts to rationalize it suggest that the stiffness should be independent of the pressure, at odds with everyday experience. Here, we study the mechanics of sheets that are a single cell thick and show how a pressure-dependent bending stiffness may arise. Our model rationalizes observations of turgor-driven shrinkage in plant cells and also suggests that turgor is unlikely to provide significant structural support in many monolayer leaves, such as those found in mosses. However, for such systems, turgor does provide a way to control leaf shape, in accordance with observations of curling upon drying of moss leaves. Guided by our results, we also present a biomimetic actuator that un-curls upon pressurization.
\end{abstract}

\maketitle

\section{Introduction}

As well as being vital for the biological function of plants generally, water plays a key role in the mechanics of herbaceous plants in particular. As examples, consider  \emph{Mimosa pudica} leaves, which are known to curl due to a change in thickness across the leaf hinge (pulvinus) caused by the motion of water between neighbouring cells \cite{Kwan2013,Sleboda2022}. Additionally, the movement of water inside bulliform cells can open and close the leaf halves of grasses in the absence of a leaf hinge \cite{Mader2020}. Water can also be used to propagate information about deformation over long distances within the plant \cite{Louf2017}. Indeed, water is a primary driver of the motion of plants \cite{Dumais2012}.

Perhaps the most obvious manifestation of water (or the lack of it) on plants is wilting under water stress. The everyday experience of plants drooping under their weight when deprived of sufficient water shows that, for many herbaceous plants, turgor must play a significant role in providing the rigidity that allows them to support their own weight. As well as being a consequence of the loss of turgor, wilting may feed back on turgor by reducing the rate of evapotranspiration and water loss \cite{Scoffoni2014}.

However, not all plants respond to the loss of water in this same way: for example, mosses, resurrection plants and reeds curl up rather than wilting \cite{Glime2017,Rafsanjani2015,Balachandran2024}, while palm leaves may exhibit corrugated folding \cite{Guo2024}. The lack of observations of wilting in mosses and other bryophytes might be due to the small size (and, hence, negligible effect of gravity) of the leaves, but nevertheless raises the question of what is the most important mechanical effect of turgor in such systems? 

The leaves of mosses have a number of unusual features compared to the leaves of larger species. Firstly, mosses are bryophytes and thus lack true vasculature; as well as having significance for the transport of water throughout the leaf, this lack of vasculature means that, except for the midrib of the leaf, there are no specific structures that stiffen it. An important second difference is that the leaves of mosses are often a single cell thick \cite{Roberts2012}. It is thus not clear \emph{a priori} how such leaves resist bending under their own weight, for example.

\begin{figure}[ht!]
\centering
\includegraphics[width=\linewidth]{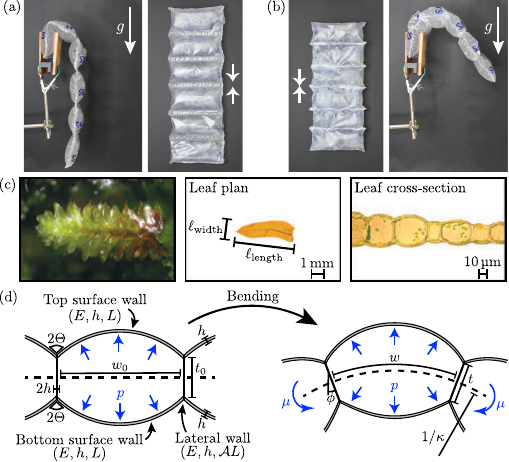}
\caption{(a,b) Gravity-driven deflections of a chain of inflated plastic `air pillows' clamped vertically at one end. In (a), the air pillows are connected by a thin flexible strip; whilst in (b), they have been glued directly together (as indicated by the white arrows). (c) Photographs of the canopy, leaf, and mid-leaf cross-section of the ground-dwelling moss \emph{Distichophyllum freycinettii}. (Images taken from \cite{Waite2010}, with permission from Wiley.) (d) Sketch of the symmetric pressurized cell model considered in \S\ref{sec:symmetriccell}. The monolayer cell sheet is bent by the application of a bending moment, $\mu$. The structure's resistance to bending is described by the effective bending modulus of the cell, $\Beff$, which we determine in terms of the cell geometry, the Young's modulus of the cell walls, $E$, and the internal pressure, $p$.\label{fig:symmetric_model}}
\end{figure}

It is natural to seek an understanding of the source of the rigidity of a monolayer leaf by seeking an analogy with highly pressurized membrane shells (such as a cylindrical balloon) \cite{Haseganu1994,Qiu2021}. As for an elastic beam, the resistance to bending of such a structure comes from the asymmetry between the in-plane stress across the cross-section of the shell. Counter-intuitively, however, this resistance to bending is independent of the internal pressure, at least for small bending deflections \cite{Haseganu1994,Qiu2021} --- pressure dependence enters only once the shell wrinkles or is subject to the Brazier instability \cite{Qiu2021}. (We give a brief explanation of this surprising result in Appendix \ref{app:Independence}.) This expectation that pneumatic pressure does not ordinarily supply significant additional mechanical stiffness to an object was confirmed by experiments \cite{SiefertThesis2019}. Nevertheless, there has been considerable interest in the use of pneumatic pressure (analogous to turgor) to control the shape of artificial structures: by designing the channel structure carefully, pneumatically-controlled structures (or `baromorphs') can be made to have an (almost) arbitrary shape in three-dimensions (3D) when actuated by pneumatic pressure \cite{Siefert2019,Gao2023}, while inflatable structures offer a readily transportable alternative to traditional structures \cite{Barton2016}. Taking more direct inspiration from biology, other solutions have used osmotically-driven flow as an actuator \cite{Sinibaldi2013,Sinibaldi2014}, or indeed plant leaves themselves \cite{Balachandran2024}. These structures often consist of inflated `tubes' connected by a thin flexible strip. These strips provide a small resistance against bending that is also independent of the tube inflation pressure \cite{Siefert2019}. Connecting the tubes directly, however, would yield a greater bending resistance dependent on the inflation pressure; this is demonstrated in Fig.~\myref{fig:symmetric_model}{a,b} using a chain of inflated `air pillows'. These connected structures are also a better reflection of the cellular structure of moss leaves, see Fig.~\myref{fig:symmetric_model}{c}. They are, thus, the main subject of interest in this paper.

This study is motivated by the twin questions of whether turgor pressure provides significant stiffening against the effects of gravity for mosses and, further, whether the loss of turgor plays a role in the curling of desiccated leaves that is observed \cite{Glime2017,Charron2009, Waite2010, Scoffoni2014, Hu2016}. To address these questions, we develop generic two-dimensional models of a single layer of pressurized cells and study their mechanical properties as well as the change of shape that occurs with changes in pressure. Since three-dimensional deformation changes the Gaussian curvature of a thin object, they are energetically expensive in comparison to two-dimensional deformations \cite{Vella2019}; as such, we anticipate that considering a two-dimensional structure will provide a useful starting point to address these questions. Moreover,  our model goes beyond the geometrical models of \citet{Gao2023} to incorporate the mechanics of the cell wall in a simple way. 

\section{Effects of pressure in  symmetric cells}\label{sec:symmetriccell}

We begin by considering the simplest model with a periodically repeated two-dimensional cell that is up-down symmetric. We shall consider the cell deformation that is induced by pressurization alone, and also study the response of a pressurized element to an imposed bending deformation. (Throughout this paper we shall assume left--right symmetry.) We consider the effect of pressure, which results from water transport in plants, but do not consider other mechanical effects of the water; in particular the bulk modulus of water is neglected.

\subsection{Model setup}\label{sec:symmetriccell_setup}

We model the two-dimensional cell as being made up of four distinct beams; these represent two surface walls (of thickness $h$, Young's modulus $E$, and relaxed length $L$), which are clamped at an exterior angle $\Theta$ to the ends of two lateral walls, see Fig.~\myref{fig:symmetric_model}{d}. (Initially, we assume the two surface walls have the same properties, but shall revisit this assumption in due course.) The two lateral walls have the same thickness and Young's modulus, but have a relaxed length $\A L$, so that the cell `aspect ratio' is $\A$; the deformed length is $t$. (Note that, because of the two-dimensional nature of the problem, the presence of natural curvature is not important in what follows.)

The cell interior is pressurized to a constant pressure $p$ and subject to a moment $\mu$, which stretches and rotates the lateral walls and deforms the surface walls. For zero applied moment, $\mu=0$, we expect the lateral walls to be vertical thanks to the up--down symmetry of the system. With non-zero $\mu$, the lateral walls are held flat by the adjacent cells, but rotate to an angle $\phi$ to the vertical at an average horizontal distance $w$ away from each other, see Fig.~\myref{fig:symmetric_model}{d}. Here, the horizontal cell width $w$, lateral wall length $t$, and angle $\phi$ are set by requiring that each cell be in equilibrium.

For the cell to be in equilibrium with the internal pressure, $p$, and bending moment, $\mu$, the lateral and surface walls must exert a force on each other, which we denote $(\TTB,\NTB)$ for the top (`$+$') and bottom (`$-$')  walls. A force and torque balance yields expressions for $\TTB$ and $\NTB$ --- see Eq.~(S1) of Ref.~\cite{SuppMat}. With these expressions, we turn to consider the deformation of each wall individually.

\subsubsection{Deformation of the lateral walls}

By construction, the lateral walls remain flat for all interior pressures and only deform longitudinally due to a constant tension (thickness-integrated stress), $\tilde{T}\coloneqq \pm\NTB \cos\phi\mp\TTB\sin\phi$. Assuming the lateral walls deform according to linear elasticity, the deformed length, $t$, satisfies Hooke's law, $\tilde{T}=Y[t/(\A L)-1]$ where $Y\coloneqq Eh$ is the two-dimensional Young's modulus of the cell walls. We, thus, find that $t$ satisfies the quadratic equation
\begin{equation}\label{eq:t}
\frac{t}{\A L} = 1+\frac{1}{Y}\left(\frac{p w}{2\cos\phi}- \frac{\mu \tan\phi}{t} \right).
\end{equation}
To determine the lateral wall angle $\phi$ and cell width $w$ requires us to consider the deformation of the surface walls.

\subsubsection{Deformation of the surface walls}\label{sec:equillateral}

We are particularly interested in turgid cells, which correspond to large interior pressures (in a sense to be defined later). This means we cannot, in general, assume the surface walls deform with small slopes. Instead, we model wall deflections using the Kirchhoff beam equations \cite{Howell2008,OReilly2017}, which account for large (nonlinear) slopes, whilst still assuming infinitesimal midline strains. 

For a Kirchhoff beam, one must distinguish between the reference (stress-free) configuration, with arclength parameter $0\leq S\leq L$, and the deformed configuration, with arclength parameter $s$. To simplify the following formulation, we work only in the reference coordinate, $S$, and present the formulation for the top and bottom walls simultaneously. This formulation follows that by  \citet{Pandey2014}.

The deformation of the top (`$+$') and bottom (`$-$') walls are measured by the axial stretches $\alpha_\pm (S)\coloneqq \de s_\pm /\de S$ and the wall positions $(x_\pm , z_\pm)$, which are given by the geometric conditions
\begin{subequations}\label{eq:Kirchhoffgeom}
\begin{equation}
\frac{\de x_\pm }{\de S} = \alpha_\pm  \cos \theta_\pm \quad \text{and} \quad \frac{\de z_\pm }{\de S} = \alpha_\pm  \sin \theta_\pm , \subtag{a,b}
\end{equation}
\end{subequations}
where $\theta_\pm (S)$ are the angle of the wall midlines with respect to the horizontal. 

Assuming the resultant moment of each wall is linearly related to the change in curvature, the intrinsic equation of the surface walls, $\theta_\pm (S)$, satisfies the Kirchhoff rod equation \cite{OReilly2017} modified to incorporate an internal pressure \cite{Marthelot2017}, which reads
\begin{equation}\label{eq:Kirchofffinal}
\begin{split}
\frac{B}{\alpha_\pm }\frac{\de^2 \theta_\pm }{\de S^2} &=\TTB \sin\theta_\pm - \NTB \cos\theta_\pm \\
&\quad\pm p\left(x_\pm \cos\theta_\pm +z_\pm \sin\theta_\pm \right),
\end{split}
\end{equation}
for $0\leq S\leq L$, where $B\coloneqq Eh^3/12$ is the bending modulus of the surface walls. Similarly, assuming the resultant tension is given by Hooke's law (linear elasticity) then the axial stretch, $\alpha_\pm (S)$, is given by
\begin{equation}\label{eq:alphafinal}
\begin{split}
Y(\alpha_\pm -1) &=\TTB \cos\theta_\pm +\NTB\sin\theta_\pm  \\
&\quad\pm p \left(z_\pm  \cos \theta_\pm  -x_\pm  \sin \theta_\pm \right).
\end{split}
\end{equation}
Note that the limit $Y\to\infty$ corresponds to an inextensible beam, for which $\alpha_\pm(S) \equiv 1$. In this limit, \eqref{eq:Kirchofffinal} recovers the equation for an \emph{elastica} subject to a pressure $p$ \cite{Howell2008, Marthelot2017}. (Detailed derivations of \eqref{eq:Kirchofffinal} and \eqref{eq:alphafinal} can be found in \S I.A.2 of \cite{SuppMat}.)

\subsubsection{Boundary conditions}

It is not clear what mechanics should be imposed at the junction between the lateral and surface walls to model the complex structure found in \eg~plant cells. For simplicity and versatility, we concentrate on clamped boundary conditions with an imposed exterior angle $0\leq\Theta\leq\pi/2$, which reflects the values seen in nature (see Fig.~\myref{fig:symmetric_model}{c}, for example) \cite{Waite2010,Hofhuis2016}. We, therefore, have the boundary conditions
\begin{subequations}\label{eq:lateralbcs}
\begin{gather}
x_\pm (0)=0,\enspace z_\pm (0)=0, \enspace x_\pm (L/2) = \frac{w}{2} \pm \frac{t}{2}\sin\phi \subtag{a--c}\\
\theta_\pm (0) =\pm\left(\frac{\pi}{2}-\Theta\right)+\phi,  \quad \theta_\pm (L/2) = 0, \subtag{d,e}
\end{gather}
\end{subequations}
where the $\pm$ refers to the top and bottom surface walls, respectively. Here, \myeqref{eq:lateralbcs}{a--c} are conditions based on the geometry of the cell, \myeqref{eq:lateralbcs}{d} is a clamping condition, and \myeqref{eq:lateralbcs}{e} comes from the symmetry of the cell.

The two extra boundary conditions for $x_\pm$, \ie~\myeqref{eq:lateralbcs}{c}, determine the unknown cell width, $w$, and lateral wall angle, $\phi$. Thus, solving \eqref{eq:Kirchhoffgeom} and \eqref{eq:Kirchofffinal} subject to \eqref{eq:lateralbcs} for both surface walls determines $w$, $\phi$, and, hence, $t$ by \eqref{eq:t}.

\subsubsection{Non-dimensionalization}\label{sec:equilb_dimensionless}

To develop an understanding of the general behaviour of our model beyond specific parameter values, we non-dimensionalize the governing system \eqref{eq:t}--\eqref{eq:lateralbcs}. In this non-dimensionalization, we use the relaxed length of the surface walls, $L$, as the reference length. This yields a system of ODEs for the surface wall angles of inclination $\theta_\pm$, axial stretches $\alpha_\pm$, and dimensionless position of a point on the walls $(X_\pm,Z_\pm)$ --- Eqs.~(S9)--(S12) of Ref.~\cite{SuppMat}. 

The non-dimensionalization just described introduces two dimensionless parameters
\begin{subequations}\label{eq:symmetricvars}
\begin{equation}
\hatp \coloneqq \frac{p L}{Y} \quad \text{and} \quad
\hatb\coloneqq \frac{B}{YL^2}, 
\subtag{a,b}
\end{equation}
\end{subequations}
which measure the internal pressure compared to the stretching of the surface walls and the relative importance of the walls' bending and stretching moduli (\ie~$B = E h^3/12$ and $Y=  E h$), respectively. ($\hatb$ may also be thought of as an inverse von K\'{a}rm\'{a}n number \cite{Plummer2020}.) The resulting system is governed by the aspect ratio of the cell, $\A$; the inverse F\"oppl--von K\'arm\'an number, $\hatb$; the dimensionless pressure, $\hatp$; and the dimensionless applied bending moment, $\hatmu\coloneqq \mu/(Y\A L)$. Using typical values for the cell wall size $L\approx\SI{50}{\micro m}$ and $h\approx\SI{1}{\micro m}$ \cite{Waite2010}, internal pressure $p\approx\SI{1}{\mega Pa}$ \cite{Proctor2007}, and Young's modulus $E\approx\SI{500}{\mega Pa}$ \cite{Proctor2007}, we find that $\hatb\sim 10^{-4}$ and $\hatp\sim10^{-1}$. Similarly, we find that   $\hatb\approx 1.2\times10^{-5}$ and $\hatp\approx 0.06$ for the inflated pneumatic devices considered by  \citet{Gao2023}. We shall thus restrict our study to relatively small pressures, $\hatp\ll 1$, which ensures small axial strains, as required for our assumption of Hooke's law. Crucially, however, we note that the cell becomes turgid when $\hatp\gg \hatb$ (pressure dominates bending), which can still be achieved with $\hatp\ll 1$ provided $\hatb\ll 1$. (To help the reader interpret the validity of results as $\hatp\to1$ appropriately, we shade plots of our numerical results to indicate the maximum strain within the beams, $\epsilon_{\max}=\max_{S}|\alpha_\pm-1|$.)

The resulting system  is  solved numerically for given dimensionless parameters $\hatmu$, $\hatp$, $\A$, and $\hatb$ using the BVP solver \texttt{bvp4c} in \texttt{\textsc{Matlab}}. This numerical solution determines the unknown lateral wall angle $\phi$, the dimensionless lateral wall length $\hatt\coloneqq t/(\A L)$, and the dimensionless  cell width $\hatw\coloneqq w/L$. The beams considered here may, in principle, have two stable states (natural and inverted  \cite{Wang2024}). However, the inverted state was not observed in our numerics: the  beams were  assumed to initially be  in the (energetically favoured) natural state and no snap-through buckling was observed as the internal pressure and bending moments varied. (Had the initial state been the inverted state, snap-through to the natural state may well have occurred.) To understand the behaviour of this system, we begin by considering the pressurized structure without any additional forcing, \ie~$\hatmu=0$.

\subsection{The pressurized configuration}\label{sec:pressurized_config}

 With no applied moment, $\hatmu=0$, the cell walls are in equilibrium with the internal pressure. The cell is up--down symmetric and we expect the lateral walls to be vertical, $\phi=0$, by symmetry. We denote the (\emph{a priori} unknown) dimensionless cell thickness and width  with no applied moment using subscript `$0$', \ie~$\hatt_0$ and $\hatw_0$.

 \subsubsection{Pressure increases cellular area}
 
\begin{figure}[ht!]
\centering
\includegraphics[width=\linewidth]{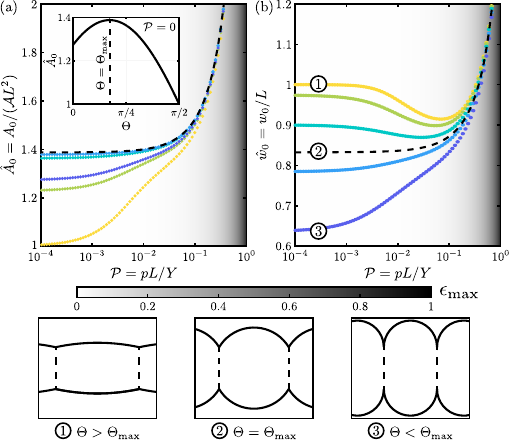}
\caption{Plots of (a) the enclosed area of the cell, $\hatA_0$, and (b) the width of the cell, $\hatw_0$, as a function of pressure, $\hatp$. These plots were numerically-determined by solving the Kirchhoff equations with parameters $\A=1/2$, $\hatb=10^{-4}$, and $\Theta=\pi/2$, $3\pi/8$, $\pi/4$, $\pi/8$, and $0$ (yellow to blue dots).  The greyscale background corresponds to the maximum strain in the top and bottom beams, $\epsilon_{\max}=\max_{S}|\alpha_\pm-1|$, as delineated by the colourbar below these plots. In the inset of (a), the area of the flaccid cell (\ie~$\hatA_0$ at $\hatp=0$) is plotted as a function of the clamping angle, $\Theta$; this is maximized at $\Theta=\Thetam\approx 0.54$, for $\A=1/2$. In the main figures, the solution for $\Theta=\Thetam$ is plotted as a black dashed curve. In this critical case, the cell is circular in shape, which qualitatively separates the different behaviours: for $\Theta>\Thetam$, the cell resembles a horizontal ellipse and an increase in pressure initially decreases the cell width, $\hatw_0$; whilst for $\Theta<\Thetam$, the cell resembles a vertical ellipse and an increase in pressure initially increases the cell width, $\hatw_0$. These regimes are delineated by \ding{172}--\ding{174}. Note that all the curves (areas and widths) converge when the dimensionless pressure becomes close to one.\label{fig:heightarea}}
\end{figure}

For a flaccid cell with no internal pressure ($\hatp=0$), the two surface walls are circular segments attached to a rectangular centre (bounded by the lateral walls). Our numerical solutions (see Fig.~\myref{fig:heightarea}{a}, for example) show that increasing the pressure, $\hatp$, increases the enclosed area, $\hatA_0$, by bending the surface walls until the cell resembles a circle with two removed segments --- corresponding to the two lateral walls --- provided that $\hatb \ll\hatp$ (\ie~the cell is turgid). (Further details are given in \S I.B of Ref.~\cite{SuppMat}, where we also show that the resistance to expansion initially increases with pressure, but ultimately decreases at large pressures. This increase in resistance could explain observations of a nonlinear volume--pressure response in plants without material nonlinear elasticity, \eg~\cite{Kierzkowski2012}.)

\subsubsection{Pressurized cells can shrink laterally}

Our numerical results show that, although turgor pressure drives an increase in cell area, it can cause cells to shrink laterally with increasing pressure, see Fig.~\myref{fig:heightarea}{b}. To understand this, apparently counter-intuitive, result, note that for large $\Theta$ the flaccid cell resembles an ellipse with horizontal major axis, and the initial increase in area caused by pressure is achieved by pulling the two lateral walls inwards ($\hatw_0$ decreases); for smaller $\Theta$, the flaccid cell resembles an ellipse with vertical major axis, and the area is initially increased by pushing out the two lateral walls ($\hatw_0$ increases), more in accordance with intuition. These two behaviours are separated by a critical inclination angle, $\Theta=\Thetam$, for which the flaccid cell is circular and the area can only be increased through stretching ($\hatw_0$ is initially constant). Example profiles of the cells in these three cases are plotted in Fig.~\myrefnb{fig:heightarea}{\ding{172}--\ding{174}}. The critical angle separating these two behaviours, $\Thetam$, is found by elementary geometry to satisfy
\begin{equation}\label{eq:thetam}
\sin\Thetam=\A\left(\pi/2-\Thetam\right).
\end{equation}

We  also note that the internal pressure affects the response of the monolayer to stretching and compression. Quantitative measures for the resistance to increasing pressure (\ie~the effective bulk modulus, $\Keff$) and to axial stretching at fixed pressure (\ie~the effective stretching modulus, $\Yeff$) can be found in \S II.A--B of Ref.~\cite{SuppMat}. We now turn to consider the effect of a non-zero applied torque, $\hatmu\neq0$.

\subsection{Response to bending: an effective bending modulus} \label{sec:Beff}

When a bending moment, $\hatmu$, is applied about the midline, the pressurized configuration is expected to bend, so that its midline adopts some curvature, $\kappa$ --- see Fig.~\myref{fig:symmetric_model}{d}. This induced curvature deforms the surface walls, breaking the up--down symmetry, and so $\phi$ is non-zero with $\kappa=2\sin\phi/w$ (by elementary geometry). The symmetry under $\hatmu\to-\hatmu$ suggests that the induced rotation $\phi\propto \hatmu$, at least for sufficiently small moments ($\hatmu\ll1$). The analogy with the linear relationship between moment and curvature of a naturally flat beam leads us to introduce an \emph{effective bending modulus}, $\Beff \coloneqq \lim_{\phi\to 0}\mu/\kappa=\lim_{\phi\to 0} \mu w/(2\phi)$.

For $\phi\ll1$, our system shows that indeed $\phi\propto \hatmu$ and the effective bending stiffness, $\Beff$, is well-defined. To non-dimensionalize $\Beff$, we use the stiffness of a flaccid cell ($\hatp=0$) with $\Theta=\pi/2$, \footnote{A more natural choice of reference state might be the flaccid cell ($\hatp=0$) with the relevant clamping angle, $\Theta$. Our choice leads to a solution independent of $\Theta$ at large pressures ($\hatp\gg \hatb$), but comes with the caveat that the effects of $\Theta$ dominate at small pressures.}. Bending this configuration induces a strain $\varepsilon=\pm \A L\kappa/2$ in the top (`$+$') and bottom (`$-$') walls, hence the effective bending modulus is $\Beff^0 \coloneqq Y(\A L)^2 /2$, and we introduce $\hatBeff\coloneqq \Beff/\Beff^0\sim\hatw_0 \hatmu/(\A\phi)$.

In general, $\hatBeff$ must be determined numerically; however, analytical progress can be made in the asymptotic limits of small and large pressures (\ie~turgid and flaccid cells). Below, we present the final asymptotic forms of $\hatBeff$; derivations can be found in  \S II.C of Ref.~\cite{SuppMat}.

\subsubsection{Flaccid cells}

For a vanishing internal pressure, $\hatp=0$, we find that $\hatBeff$ is determined by the geometry of the cell, \ie~$\hatBeff\sim \hatBeff^0(\Theta,\A,\hatb)$ for $\hatp\ll1$ where $\hatBeff^0$ is (S31) of Ref.~\cite{SuppMat}. Here, the dependence on $\hatb\ll 1$ determines whether the surface walls respond by bending or stretching and $\A$ controls the aspect ratio of the cell. For $\Theta=\pi/2$, $\hatBeff^0=1$ (by choice of reference value), whilst for $\Theta = 0$, $\hatBeff^0 = 4\pi(\pi^2+4/\A)\hatb/(\pi^4\hatb+\pi^2-8) \ll 1$; this agrees with the observation that $\hatBeff$ increases with the exterior clamping angle, $\Theta$, see Fig.~\ref{fig:hatBeff_p}.

\subsubsection{Turgid cells}

For a large internal pressure, $\hatp\gg \hatb$, we find that 
\begin{equation}\label{eq:Beff_largep}
B_\mathrm{eff} \sim \frac{2 R_0^3 t_0w_0 }{2R_0 w_0-Lt_0}p,
\end{equation}
where $R_0\coloneqq (w_0^2+t_0^2)^{1/2}/2$, $w_0$, and $t_0$ are the pressurized cell's radius, width, and thickness, respectively, which also depend on the pressure. For $\hatb\ll\hatp\ll1$, \eqref{eq:Beff_largep} increases linearly with the pressure. For moderate pressures, $\hatp\sim1$, we see a superlinear effect, although at such values the straining of the beams is important and so, the assumption of small strains may no longer be  valid (hence the shading in Fig.~\ref{fig:hatBeff_p}). An angular stiffness that is linear in the pressure has previously been found for turgid pressurized cells \cite{Gao2023}. However, the effects of wall stretching, which gives rise to the observed superlinear response, has not, to our knowledge, been appreciated previously.

\subsubsection{Moderate internal pressures}

\begin{figure}
\centering
\includegraphics[width=\linewidth]{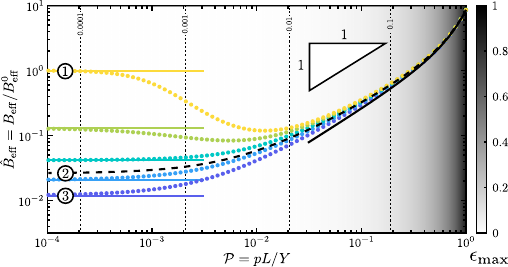}
\caption{Plot of the effective bending modulus, $\hatBeff$, as a function of the internal pressure, $\hatp$, for the configuration parameters $\A=1/2$, $\hatb=10^{-4}$, and $\Theta=\pi/2$, $3\pi/8$, $\pi/4$, $\pi/8$, and $0$ (yellow to blue). The effective moduli found by numerically solving the Kirchhoff equations are plotted as coloured dots, while asymptotic results are plotted as solid curves: $\hatBeff\sim\hatBeff^0$, valid for small pressures, is shown with the corresponding colour and \eqref{eq:Beff_largep}, valid for large pressures, is shown in black. The black dashed curve is the numerical solution with maximum encapsulated area, $\Theta=\Thetam\approx 0.54$. The greyscale background corresponds to the maximum strain in the top and bottom beams, $\epsilon_{\max}=\max_{S}|\alpha_\pm-1|$,  as delineated by the colourbar; the contour lines of $\epsilon_{\max}=10^{-4}$, $10^{-3}$, $10^{-2}$, and $10^{-1}$ are shown as vertical dashed lines. \ding{172}--\ding{174} correspond to the three regimes sketched in Fig.~\ref{fig:heightarea}. Although $\hatBeff$ may initially decrease with $\hatp$ (\ie~for large values of $\Theta$), it ultimately increases with $\hatp$. Note that all the curves (bending stiffnesses) converge when the dimensionless pressure becomes close to one.\label{fig:hatBeff_p}}
\end{figure}

In Fig.~\ref{fig:hatBeff_p}, the numerically determined $\hatBeff$ is plotted as a function of $\hatp$ for $\A=1/2$, $\hatb=10^{-4}$, and various $\Theta$. Also plotted are the asymptotic results at small pressures, \ie~$\hatBeff\sim \hatBeff^0$, and large pressures, \ie~\eqref{eq:Beff_largep}. As well as showing that the expected asymptotic behaviour is recovered by these numerics, Fig.~\ref{fig:hatBeff_p} shows that $\hatBeff$ may initially decrease with pressure, $\hatp$, if the exterior clamping angle, $\Theta$, is sufficiently large, but ultimately it increases with $\hatp$ according to \eqref{eq:Beff_largep}. Altering the configuration parameters, $\A$ and $\hatb$, changes the results presented here quantitatively, but not qualitatively.

\section{Pressure-dependent morphing in asymmetric cells}
\label{sec:asymmetric}

Many leaves are observed to curve when they dry \cite{Glime2017,Rafsanjani2015,Guo2024}. One benefit of curving a water-stressed leaf is to decrease the rate of evaporation, while the benefit of being flat when not water-stressed is to increase the area available to intercept sunlight \cite{Guo2024}. However, an up--down symmetric structure will not change its moment-free curvature, regardless of the turgor pressure. To understand the turgor-driven curvature of leaves, therefore, requires an asymmetry, either mechanical or geometrical, to be introduced at the cellular level. In this section, we consider the effect of different asymmetries on the variation of tissue curvature with pressure.

\subsection{Equilibrium equations of the asymmetric cell}\label{sec:asymmetricform}

\begin{figure}[ht!]
\centering
\includegraphics[width=\linewidth]{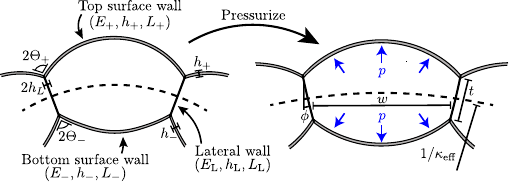}
\caption{ Sketch of the asymmetric pressurized cell model considered in this section. The top and bottom surface walls have differing values of Young's moduli $\ETB$, thicknesses $\hTB$, relaxed lengths $\LTB$, and/or clamping angles $\ThetaTB$. These asymmetries induce an effective curvature, $\kapeff$, which varies with the internal pressure, $p$. \label{fig:asymmetric_model}}
\end{figure}

To introduce an asymmetry in our cell model, we allow the surface walls to differ in one or more of the following properties: relaxed length $L$, thickness $h$, Young's modulus $E$, and/or clamping angle $\Theta$. We, therefore, replace $\{E,h,L,\Theta\}$ with $\{\ETB,\hTB,\LTB, \ThetaTB\}$ for   the top (`$+$') and bottom (`$-$') surface walls. Other than this notational change, the mathematical formulation for such an asymmetric cell is identical to the symmetric case considered in \S\ref{sec:symmetriccell}, but with $\mu=0$ since we neglect any applied bending moment. For clarity, here, we also relabel the Young's modulus, thickness, and relaxed length of the lateral walls as $\ELat$, $\hLat$, and $\LLat$, respectively. This setup is sketched in Fig.~\ref{fig:asymmetric_model}.

Using the relaxed lengths of the surface walls as the reference lengths, the surface walls' slopes $\theta_\pm$, axial stretches $\alpha_\pm$, and dimensionless position of a point on the surface walls $(X_\pm,Z_\pm)$ satisfy a system of ODEs, Eqs.~(S9)--(S13) of Ref.~\cite{SuppMat}. These equations are governed by the same dimensionless parameters that were introduced for the symmetric case (\S\ref{sec:symmetriccell_setup}), albeit with subscript distinguishing between the top (`$+$') and bottom (`$-$') surface walls, \ie~$\hatpTB$, $\ATB$, $\hatbTB$, $\ThetaTB$, and $\hatwTB$. Additionally, we introduce a new parameter:
\begin{equation}
\hatYTB \coloneqq \frac{\ELat\hLat}{\ETB \hTB},
\end{equation}
which is the ratio of the stretching moduli for the top and bottom walls compared to the lateral walls. (Note that $\hatYTB= 1$ for the homogeneous symmetric cell considered in \S\ref{sec:symmetriccell_setup}.)

By introducing the parameter ratios,
\begin{subequations}\label{eq:asymmetricpars}
\begin{align}
\RE \coloneqq \EB/\ET,&\quad \Rh \coloneqq \hB/\hT, \subtag{a,b}\\
 \RL \coloneqq \LB/\LT,&\quad \RT \coloneqq \ThetaB/\ThetaT,\subtag{c,d}
\end{align}
\end{subequations}
the parameters for the bottom  wall can be written in terms of those for the top  wall. The difference of the ratios in \eqref{eq:asymmetricpars} from unity measures the asymmetry of the cell. Overall, given values for the top surface wall parameters ($\hatpT$, $\AT$, $\hatbT$, $\hatYT$, and $\ThetaT$) and the asymmetric ratios ($\RE$, $\RL$, $\Rh$, and $\RT$), the system for the top and bottom  walls can be solved to determine the cell width $\hatwT\equiv\RL\hatwB$, cell thickness $\hatt$, and lateral wall angle $\phi$. We now turn to consider a macroscopic measure of the asymmetric cells, namely the effective curvature.

\subsection{Effective curvature}\label{sec:intrinsiccurv}

At the scale of a whole leaf, individual cells cannot be identified, instead the asymmetry in \eg~plant tissue is observed as an intrinsic curvature. In the case of the asymmetric cell model introduced in \S\ref{sec:asymmetricform}, this \emph{effective curvature} is the curvature of the cell midline which, from simple geometrical arguments, can be written as
\begin{equation}\label{eq:kapeff}
\kapeff\coloneqq 2\sin\phi/ w,
\end{equation}
where we associate a positive curvature with a positive angle $\phi$ for the left lateral wall (\ie~the cell curves downwards, Fig.~\ref{fig:asymmetric_model}). As we show below, the scale of $\kapeff$ and how it varies with $\hatp$ is dependent on the type and magnitude of asymmetry chosen; thus, there is no obvious choice of a dimensionless scale for $\kapeff$. Instead, we choose $\LLat$ (the relaxed length of the lateral walls) as a reference length to aid the comparison between the different asymmetries, \ie~we define $\hatkapeff\coloneqq \LLat\kapeff = 2\ATB\sin\phi/\hatwTB$. 

\begin{figure}[ht!]
\centering
\includegraphics[width=\linewidth]{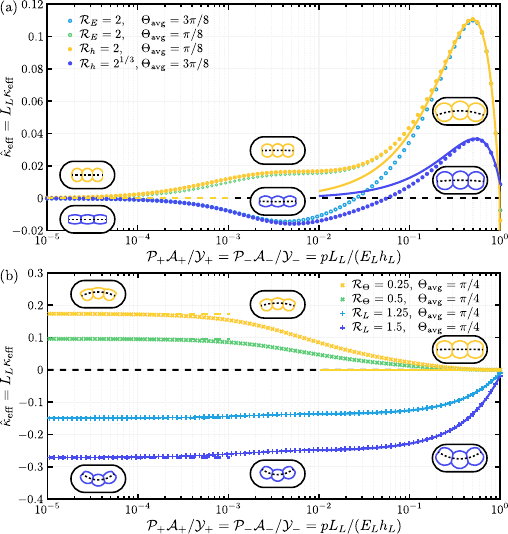}
\caption{Plots of the numerically-determined effective cell curvature, $\hatkapeff$, as a function of the internal pressure for an asymmetry in either the Young's modulus ($\figwcircle$ in panel a), thickness ($\figbcircle$ in panel a), relaxed length ($\figplus$ in panel b), or clamping angle ($\figtimes$ in panel b) of the surface walls, where the corresponding values of $\RE$, $\Rh$, $\RL$, or $\RT$ (asymmetries in the Young's modulus, thickness, length, and clamping angle, respectively) are given in the keys. (Note that the ratio values, and hence the sign of curvature, are chosen for presentation purposes only.) The parameters are chosen such that their averages recover those used in Fig.~\ref{fig:hatBeff_p}, that is $(1/\hatYT+1/\hatYB)/2 =1$, $(1/\AT+1/\AB)/2=1/2$, $(\hatbT+\hatbB)/2=10^{-4}$, and $(\ThetaT+\ThetaB)/2=\Theta_\mathrm{avg}$, where in (a),~$\Theta_\mathrm{avg}=3\pi/8$ (blues) and $\pi/8$ (yellow and green), and in (b),~$\Theta_\mathrm{avg}=\pi/4$. Also plotted are the asymptotic results for small pressures (as dashed lines) and large pressures (as solid curves), as derived in \S II.D of Ref.~\cite{SuppMat}, and the cell profiles for the asymmetry with the corresponding colour at $\hatp = 0$, $10^{-5/2}$, and $10^{-1/2}$ (left to right).\label{fig:hatkap}}
\end{figure}

In Fig.~\ref{fig:hatkap}, we plot the curvature of the cell, $\hatkapeff$, obtained numerically, as a function of the internal pressure with one asymmetry introduced at a time. Also plotted are asymptotic predictions, which hold for small and large pressures --- details of these can be found in \S II.D of Ref.~\cite{SuppMat}. It is apparent that the different types of asymmetry induce variations of curvature with pressure that are qualitatively different. Below, we explain the results for each type of asymmetry.

\paragraph*{Asymmetry in the Young's modulus $(\RE\neq 1)$} At small internal pressures, the surface walls are unstrained since their ends are free to move horizontally; thus, an asymmetry in the Young's modulus of the surface walls does not induce a curvature at small pressures. As the pressure increases, however, the walls deform in response to the additional loading. Initially, the cell curves towards or away from the more compliant wall, depending on whether decreasing or increasing the width of the cell increases its area --- \ie~$\Theta<\Thetam$ or $\Theta>\Thetam$, respectively, with $\Thetam$ given by \eqref{eq:thetam}. (In the case of Fig.~\myref{fig:hatkap}{a}, $\RE=2>1$, so the top surface wall is more compliant; thus, for $\Theta=3\pi/8> \Thetam\approx 0.54$, the cell initially bends upwards and $\hatkapeff$ is negative, whilst for $\Theta=\pi/8< \Thetam\approx 0.54$, it bends downwards and $\hatkapeff$ is positive.) As the pressure increases further, \ie~for $\hatpTB\gg\hatbTB$, the cell approaches a circular arc, thus the walls do not bend any more and, instead, stretch under an increase in pressure. Again, the more compliant wall deforms at lower pressures; this increases its length and bends the cell away from it, regardless of the clamping angle, $\Theta$. (In Fig.~\myref{fig:hatkap}{a}, the cell curves downwards at large pressures for both $\Theta=3\pi/8$ and $\pi/8$.) Finally, for $\hatpTB\sim 1$, the pressure is sufficiently large that both walls are able to stretch freely, resulting in the cell flattening.

\paragraph*{Asymmetry in the surface wall thicknesses $(\Rh\neq 1)$} An asymmetry in the surface wall thickness is analogous to the asymmetry in the Young's modulus, with the induced curvature depending on the pressure in a qualitatively identical way. In fact, since the walls only bend at small pressures and the bending modulus $B\propto E h^3$, taking $\RE=\mathcal{R}$ and $\Rh=1$ is the same as taking $\RE=1$ and $\Rh=\mathcal{R}^{1/3}$ at small pressures. Conversely, for large pressures, it is the stretching modulus $Y\propto Eh$ that matters, so taking $\RE=\mathcal{R}$ and $\Rh=1$ is the same as taking $\RE=1$ and $\Rh=\mathcal{R}$ at large pressures. This correspondence is shown in Fig.~\myref{fig:hatkap}{a} for $\mathcal{R}=2$. 

\paragraph*{Asymmetry in the surface wall lengths $(\RL\neq 1)$} If one of the surface walls is naturally longer than the other, then the cell must be curved in its natural state. (In the case of Fig.~\myref{fig:hatkap}{b}, $\RL>1$, so the bottom surface wall is longer than the top and the cell is curved upwards --- $\hatkapeff$ is negative.) As the pressure increases, the curvature smoothly transitions from the results for a flaccid cell (governed by geometry) to those of a turgid cell, see \S II.C of Ref.~\cite{SuppMat}. At very large pressures, $\hatpTB\gtrsim1$, the walls are free to stretch and so the cell grows, resulting in a smaller effective curvature. 

\paragraph*{Asymmetry in the clamping angles $(\RT\neq 1)$} For small internal pressures, an asymmetry in the clamping angles causes the cell to be curved away from the shallower-angled surface wall --- \ie~the larger value between $\ThetaT$ and $\ThetaB$. (In Fig.~\myref{fig:hatkap}{b}, $\RT<1$ so that $\ThetaB<\ThetaT$ and, hence, the cell is curved downwards with a positive $\hatkapeff$.) This behaviour is determined purely by elementary geometry. As the pressure increases, however, the importance of the clamped boundary decreases; in particular, for $\hatpTB\gg\hatbTB$ the clamping angles are irrelevant except in boundary layers of width $\Oh(\hatbTB/\hatpTB)$ close to the ends of the surface walls. Outside these boundary layers, the cell is effectively symmetric with zero curvature. Unlike the other asymmetries, this trend towards zero-curvature can occur at small axial strains ($\hatpTB\ll1$), where the assumption of linear elasticity still holds, see Fig.~\myref{fig:hatkap}{b}.

\smallskip
Overall, a geometrical asymmetry (\eg~$\RL\neq1$ or $\RT\neq1$) tends to lead to a curvature when flaccid, which vanishes as the internal pressure increases; while a mechanical asymmetry (\eg~$\RE\neq1$ or $\Rh\neq1$) tends to lead to no curvature when flaccid and a high curvature when turgid. In reality, multiple asymmetric properties might be combined in such a way as to have conflicting effects, and so the curvature, $\hatkapeff$, might vary with the pressure differently to that discussed. However, the asymptotic results for small and large pressures \cite{SuppMat}, still hold true and can be used as a first step towards understanding how $\hatkapeff$ varies with pressure.

\section{Biological implications\label{sec:Biology}}

In \S\ref{sec:symmetriccell} and \S\ref{sec:asymmetric}, we presented the results of detailed model calculations for a monolayer cell sheet subject to an internal pressurization. To understand the effective rigidity of such sheets, we focussed on cell sheets that are up-down symmetric  (\S\ref{sec:symmetriccell}); this also allowed us to understand changes in length of the sheet caused by pressurization. Cell sheets with embedded up-down asymmetry added the possibility of out-of-plane shape changes (\S\ref{sec:asymmetric}). We now turn to several biological systems that share the essential ingredients we have discussed to learn what our modelling tells us about these scenarios.

\subsection{Length changes during pressurization}

Our model predicted that whether a symmetric cell sheet shrinks or expands upon increasing pressurization depends on the angle at which the beams are clamped. If $\Theta<\Thetam$, pressure acts to `iron out' the excess length stored in the arched beams and each cell extends laterally. This behaviour is almost intuitive, and has been reported as part of the explanation for the growth in size of the wings of \emph{Drosophila melanogaster} as they emerge from the pupal stage \cite{Hadjaje2024}.

However,  if $\Theta>\Thetam$, the opposite behaviour is observed: because the beams are less inclined in this case, their primary response to pressurization is to become more highly curved, which, without stretching, requires the cell to shrink laterally. Though less intuitive, this pressure-induced shrinkage  \emph{has} been observed in plant tissue and is highly dependent on the cell geometry and wall anisotropy \cite{Hofhuis2016}. Using cellular force microscopy data, \citet{Hofhuis2016} fitted experimental data for the wall elastic moduli, obtaining $E_{\parallel}\approx \SI{9000}{\mega\pascal}$ and $E_{\perp}\approx \SI{150}{\mega\pascal}$ parallel and perpendicular to the shrinking axes, respectively. With these Young's moduli, their finite element method (FEM) simulations suggest a relative cell length $R_\mathrm{len}\approx0.91$ and volume $R_\mathrm{vol}\approx1.53$ at a turgor pressure $p\approx \SI{0.65}{\mega\pascal}$. These are comparable to the values $R_\mathrm{len}\approx 0.88$ and $R_\mathrm{vol}\approx 1.53$ observed in their experiments. Their results can be directly compared to our two-dimensional model since they assume the cell wall in the third dimension (\ie~perpendicular to the shrinking axis) is soft in comparison to the other dimensions (\ie~parallel to the shrinking axis). Using our model with the same cell parameters (\ie~$h=\SI{0.2}{\micro\metre}$, $L=\SI{50}{\micro\metre}$, $\A = 0.4$, $p\approx \SI{0.65}{\mega\pascal}$, $E=\SI{9000}{\mega\pascal}$, and $\Theta=\pi/2$), we obtain $R_\mathrm{len}=\hatw_0\approx 0.83$ and $R_\mathrm{vol}= \hatA_0\approx 1.53 $, which are remarkably similar to the values obtained by their more sophisticated FEM simulations \footnote{Note that a slightly pressurized cell was used as a reference state by Ref.~\cite{Hofhuis2016}, to reflect the observations of actual plant cells. This is in comparison to the truly unpressurized state ($\hatp=0$), that we have used here. Using a pressurized cell has the effect of slightly increasing $R_\mathrm{len}$ and slightly decreasing $R_\mathrm{vol}$, but without more details of the slight pressurization used by \citet{Hofhuis2016} we cannot quantify this effect in our results.}.

Pressure-induced shrinkage has also been used to develop bio-inspired programmable shells \cite{Siefert2019,Siefert2020, Gao2023} and, in tandem with an element of fixed length, to induce curvature \cite{Sleboda2022}. In particular, the linear increase of curvature with pressure observed experimentally there is consistent with our theoretical work, which suggests a linear increase in size with increasing pressure.

\subsection{Size limits for monolayer leaves}\label{sec:sizelimits}

The model developed thus far was motivated by the leaves of mosses, which are often a single-cell thick, and have no vasculature to stiffen them. Having developed the model of \S\ref{sec:symmetriccell}, it is natural to use it to understand whether the size of a moss leaf is limited by the ability of turgor pressure to support it against gravity.

\subsubsection{Bending under gravity}

A horizontally clamped beam (of bending modulus $B$ and linear density $\varrho$) bends under the influence of its weight when its length, $\ell$, is comparable to the elastogravitational length $\ellbg\coloneqq(B/\varrho g)^{1/3}$, with $g$ the gravitational acceleration \cite{Wang1986,Holmes2019}: if $\ell\gg \ellbg$, the beam will sag significantly. Such sagging could have two detrimental consequences for a plant leaf: (i) it decreases the incident sunlight that may be captured and (ii) it may lead to large internal strains, which may cause structural failure and damage. Previously, similar constraints for a \emph{heavy elastica} have been used to compute the branch shape with maximum reach \cite{Wei2012} and the leaf design that maximizes the light intercepted for photosynthesis \cite{Tadrist2016}. By taking $B=\Beff$ and $\varrho = \rho \ellthick$, where $\rho$ is the bulk density of the tissue and $\ellthick$ is the macroscopic thickness, we can compute the typical length scale over which our cellular structure would bend under gravity,  \ie~$\ellbg$; we then compare $\ellbg$ to the lateral dimensions of naturally occurring moss leaves, specifically the leaf length and width, $\ellwidth$ and $\elllength$, as defined in Fig.~\myref{fig:symmetric_model}{c}.\footnote{Note that both the length scale over which the beam strains due to bending, $\ellstrain\coloneqq [\Beff/(\rho g\ellthick^2)]^{1/2}$, and stretches due to gravity, $\ellstretch\coloneqq \Yeff/(\rho g\ellthick)$, where $\Yeff$ is the effective stretching modulus, are typically orders of magnitude larger than the elastogravitational length, $\ellbg$, and so are not considered here; nevertheless, further details can be found in Appendix~\ref{app:further_lim}.} 
 
For small deflections, the change in the horizontal projected area (and, hence, area intercepting sunlight) is $L-x(L)=\int_0^L(1-\cos\theta)~\de S\propto \int_0^L\theta^2~\de S$ and, thus, is expected to be quadratic in the relative deflection. The typical relative deflection of the heavy elastica is 
\begin{equation}\label{eq:mathcalDE}
\mathcal{D} \coloneqq \frac{w(\ell)}{\ell} = \frac{\ell^3}{8\ellbg^3}, 
\end{equation}
where $w(x)$, for $0\leq x\leq \ell$, is the profile of the deformed beam (see Appendix~\ref{app:further_lim} and Ref.~\cite{Howell2008}). In conclusion, sufficiently small leaves ($\ell\ll2\ellbg$) should not sag significantly under gravity. With the measure of deflection \eqref{eq:mathcalDE}, we can now test whether turgor pressure  is required for rigidifying the width and length of moss leaves against gravity. We shall begin by considering the leaf widths since, unlike the leaf lengths, these do not have the additional support provided by a leaf costa/midrib.

\subsubsection{Effect of the width of  moss leaves}

\begin{figure}[ht!]
\centering
\includegraphics[width=\linewidth]{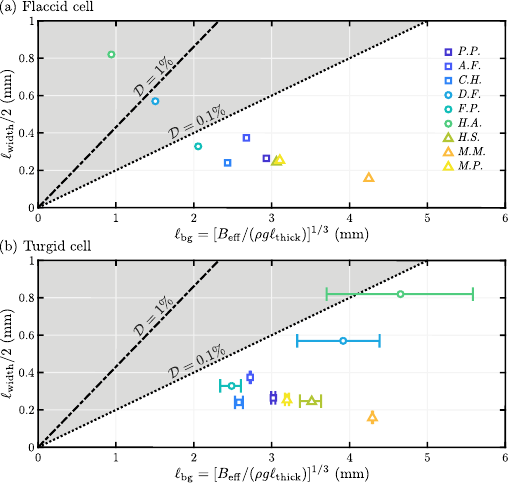}
\caption{Plot of the extended lamina widths, $\ellwidth/2$, against estimated elastogravitational length, $\ellbg=[\Beff/(\rho g \ellthick)]^{1/3}$, for nine species of single-cell thick moss; here, colour denotes the species (see key), while point shape denotes the type as follows: trunk-dwelling mosses, $\figwsquare$, (\textit{Pyrrhobryum pungens}, \textit{Acroporium fuscoflavum}, \textit{Campylopus hawaiicus}); ground-dwelling mosses, $\figwcircle$, (\textit{Distichophyllum freycinetii}, \textit{Fissidens pacificus}, \textit{Hookeria acutifolia}); branch-dwelling mosses, $\figwtriangle$, (\textit{Holomitrium seticalycinum}, \textit{Macromitrium microstomum}, \textit{Macromitrium piliferum}). Here, we have used the cell width, cell height, cell wall thicknesses, lamina thickness, and maximum leaf width reported by \citet{Waite2010}  for each species to estimate $L$, $\A L$, $h$, $\ellthick$, and $\ellwidth$, respectively. We further assume a density $\rho=\SI{1000}{\kilogram/\metre\cubed}$; clamping angle $\Theta=\pi/4$ (based on the cross-sections shown in \cite{Waite2010}); Young's moduli $E= \SI{100}{\mega\pascal}$, which is chosen as a lower bound to the typically accepted range in plant cells \cite{Cosgrove2016}; and turgor pressures (a) $p = \SI{0}{\mega\pascal}$ (\ie~a flaccid cell) and (b) $\SI{0.8}{\mega\pascal}\leq p\leq \SI{2.1}{\mega\pascal}$ (the typical range found in moss \cite{Proctor2007}, with variation shown by error bars in $\ellbg$). The lines $\ellwidth/2=2 \mathcal{D}^{1/3} \ellbg$ for $\mathcal{D} =10^{-2}$ (dash-dotted) and $\mathcal{D} =10^{-3}$ (dotted) correspond to the deflection, $\mathcal{D}$, of an Euler--Bernoulli beam, \eqref{eq:mathcalDE}; thus, the shaded region corresponds to where appreciable deformation under gravity would be expected. \label{fig:Waite}}
\end{figure}

In Fig.~\ref{fig:Waite}, the leaf widths, $\ellwidth$, of nine different species of single-cell thick moss are plotted against the corresponding elastogravitational lengths, $\ellbg= [\Beff/(\rho g\ellthick )]^{1/3}$, predicted from our model. Here, we use geometrical parameters for the moss leaves and their cells from Ref.~\cite{Waite2010} and take the Young's modulus of the cell walls as $E = \SI{100}{\mega\pascal}$ (a lower bound on the typically accepted values for plant cells \cite{Cosgrove2016}). In Fig.~\myref{fig:Waite}{a}, the prediction of $\ellbg$ is calculated assuming that the cells are flaccid ($\hatp= 0$), while in Fig.~\myref{fig:Waite}{b}, $\ellbg$ is calculated assuming a turgor pressure in the range $\SI{0.8}{\mega\pascal}\leq p\leq \SI{2.1}{\mega\pascal}$ (to reflect the typical values found in moss \cite{Proctor2007}). 

By comparing Figs.~\myref{fig:Waite}{a} and \myref{fig:Waite}{b}, it is apparent that turgor pressure increases the elastogravitational length, $\ellbg$. In fact, using the small deflection solution, \eqref{eq:mathcalDE} with $\ell= \ellwidth/2$, the expected deflection of turgid leaves are all less than $1\%$, \ie~$[(\ellwidth/2)/\ellbg]^{3}= 8\mathcal{D}<8\times10^{-2}$. Nevertheless, even for flaccid cells ($\hatp=0$), the expected deflection remains below a few percent.

\subsubsection{Effect of the length of  moss leaves}

The leaf length, $\elllength$, of the moss species studied by Ref.~\cite{Waite2010} is typically 3--17 times larger than their width, $\ellwidth$. Multiplying the abscissa of the data plotted in Fig.~\ref{fig:Waite} by such a large factor would move all data points into the `dangerous' region where the leaf would be expected to sag under its own weight. Of course, the leaf costa/midrib provides additional longitudinal support along the length of the leaf and so the simple picture used to understand the likely deformation under gravity of the width of the leaf is not strictly relevant. Nevertheless, the fact that this midrib is significantly thicker than neighbouring tissue (and, hence, does rigidify the leaf) is consistent with the result of our theory that moss leaves are relatively close to the threshold at which gravitational deformation does become significant.

Our results suggest that the widths of leaves are small enough that they do not need rigidifying, but that the lengths are long enough that: (i) turgor does not provide significant additional support beyond that intrinsic to the cell walls and so (ii) the midrib is required to provide structural support against gravitational effects.

\subsection{Curvature in plant leaves}

We have shown that turgor pressure does not provide a significant advantage in terms of rigidifying the leaf against the effect of gravity. We, therefore, turn now to consider how pressure might produce the significant changes in curvature that are seen upon the desiccation of leaves. A hydrated (large turgor pressure) moss leaf is flat to maximize light interception, while a dehydrated (small turgor pressure) moss leaf is tightly curved, which limits further water loss \cite{Charron2009,Hu2016,Guo2024}.

\begin{figure}[ht!]
\centering
\includegraphics[width=\linewidth]{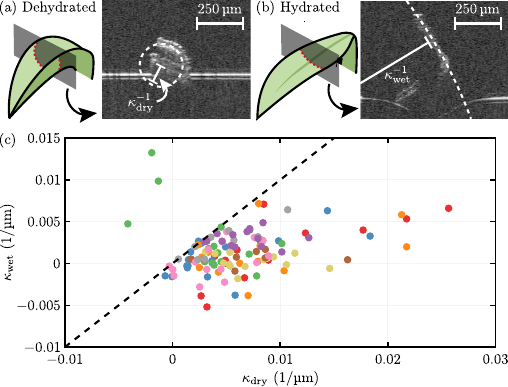}
\caption{Optical Coherence Tomography images of a cross-section of a (a) dehydrated and (b) hydrated \emph{Physcomitrium patens} moss leaf. The two laminar halves protruding from the midrib are approximated by a circle (shown as dashed curves), the radius of which yields the dry and wet signed curvatures, $\kappa_\mathrm{dry}$ and $\kappa_\mathrm{wet}$, respectively. (c) Plot of $\kappa_\mathrm{wet}$ against $\kappa_\mathrm{dry}$ for $11$ \emph{Physcomitrium patens} leaves at various cross-sections along the leaves' lengths (points with the same colour correspond to measurements at different points along the same leaf). The majority of the points lie below $y=x$ (black dashed line), showing the leaves tend to curl when dehydrated.\label{fig:MossExpts}}
\end{figure}

To illustrate more concretely this change in curvature, Fig.~\myref{fig:MossExpts}{a,b} shows Optical Coherence Tomography images of \emph{Physcomitrium patens} moss leaves when hydrated (turgid) and dehydrated (flaccid). By taking cross-sections of the leaves at fixed points along their length, the effective curvature of the two lamina halves protruding from the midrib were measured by fitting a circle, as demonstrated in Fig.~\myref{fig:MossExpts}{a,b}. (Further details of these  experiments can be found in Appendix \ref{app:experiment}.) Fig.~\myref{fig:MossExpts}{c} shows a plot of the hydrated curvature against the dehydrated curvature for a number of leaves and cross-sections. The majority of these points lie below the line $y=x$, suggesting the leaves do indeed curl when they dry. A linear least-squares fit yields a gradient of $ 0.2169$ with a coefficient of determination of $R^2 = 0.2075$ after omitting the three outliers above the $y=x$ line.

We  studied the pressure dependence of the curvatures induced by the various asymmetries in Fig.~\ref{fig:hatkap}. From this, it appears that an asymmetry in the clamping angles ($\RT\neq 1$) is closest to qualitatively reproducing the  observation that a dehydrated leaf is curved, whilst a hydrated leaf is flat.  The green and yellow curves of fig.~\myref{fig:hatkap}{b} show a curved state at low pressure (dehydrated) that is gradually ironed out by bending as the pressure increases (hydration). Figure \myref{fig:hatkap}{b} shows that an asymmetry in the length of the surface walls also qualitatively recovers this behaviour. However, this is only achieved with $\hatp\approx1$ when $\RL\neq1$ (see blue curves in fig.~\myref{fig:hatkap}{b}) because in this case the walls need to stretch (rather than bend) to flatten the leaf.

The results above raise the question of why the curvature control mechanism proposed above is not used by vascular plants. Indeed, resurrection plants seem to curl up by relying on gradients in density of lignin, a hydrophobic component of the cell wall \cite{Rafsanjani2015}. Most vascular plants wilt when they loose turgor. This difference can be rationalized as follows. In contrast to a curvature induced by dehydration, turgid plant tissue has also been observed to curl; for example, longitudinal strips removed from \emph{Tulipa} stems spontaneously recurve \cite{Niklas1997}; detached exocarp cells after explosive seed dispersal in \emph{Cardamine hirsuta} curl away from the fruit \cite{Hofhuis2016}; and animal epithelial monolayers curl downwards when devoid of a supporting substrate \cite{Fouchard2020}. In all three of these examples it was proposed that the spontaneous curvature was due to an anisotropy between the outer and inner layers of the cellular structure: \citet{Niklas1997} suggested that the spontaneous curvature was due to differing mechanical properties across the cell walls (as first suggested by \citet{Hofmeister1859}); \citet{Hofhuis2016} modelled the curling of the detached exocarp as the release of tension inside a structure made of three different materials: an active soft outer layer, a passive middle, and a stiff inner layer; and \citet{Fouchard2020} proposed that the asymmetry of the animal cell sheet was due to an enrichment of Myosin II in the substrate-supported side. Parallels can, thus, be drawn between these observations and the asymmetric Young's modulus (or equivalently thickness) induced curvature, $\RE\neq 1$ (or $\Rh\neq 1$). Accordingly, our model provides a tool for the qualitative understanding of the curvature of simple cellular structures.

\section{Biomimetic control of curvature}

In addition to the biological examples considered in \S\ref{sec:Biology}, the pressure-sensitive curvature of the monolayer cellular sheets considered here may open new possibilities for the design of inflatable structures \cite{Cadogan2004,Siefert2020,Gao2023}. Typically, such structures are  flat in the unpressurized state and curve with increasing pressure \cite{Siefert2020,Jones2021,Chen2024}; this is usually caused by asymmetries in the thickness or modulus of different sides of the object (see, for example, Ref.~\cite{Jones2021})  and hence gives curvature--pressure evolution of the type shown in Fig.~\myref{fig:hatkap}{a}. However, Fig.~\myref{fig:hatkap}{b} shows that asymmetry in the angles or lengths of the cells may lead to a very different behaviour with the system is initially curved but becoming flat as the pressure increases. As a concrete example, one can imagine that this curvature--pressure evolution might find application as a passive soft robotic gripper, with the unusual feature that the unpressurized device is clenched (\ie~curved), whilst the pressurized device is open (\ie~flat) \cite{Ilievski2011,Rus2015}. 

\begin{figure}[ht!]
\centering
\includegraphics[width=\linewidth]{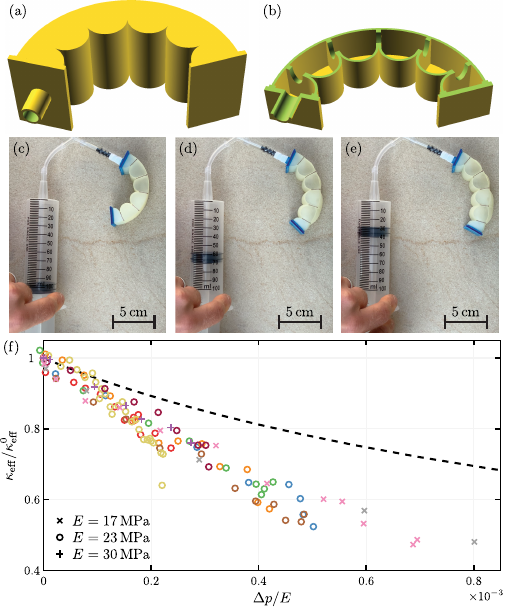}
\caption{A 3D-printed realization of the asymmetric pressurized structure of Fig.~\ref{fig:asymmetric_model}  uncurves upon pressurization. Panel (a)  shows a 3D render of the  full computer aided design (CAD), whilst panel (b) shows a half section. To enable inflation, the device has been designed with a `lid' and a `base' in the third-dimension (separated by $\SI{30}{\milli\metre}$). The other geometrical parameters are $\LLat= \SI{10}{\milli\metre}$, $\LTB=\SI{20}{\milli\metre}$, $\hLat=\SI{1}{\milli\metre}$, $\hTB=\SI{1}{\milli\metre}$, $\ThetaT = 0$, and $\ThetaB=\pi/2$. Equal pressurization of each cell is ensured by elliptical holes (of width $\SI{6}{\milli\metre}$ and length $\SI{12}{\milli\metre}$) that connect neighbouring cells --- as seen in (b). The structure was 3D printed using the PolyJet elastomer Agilus-30 \cite{Abayazid2020}. Panels (d)--(e) show increasing pressurizations ($\Delta p =0$, $\SI{5.87}{\kilo\pascal}$ and $\SI{8.13}{\kilo\pascal}$, respectively) of the 3D-printed structure with the effective curvature, $\kapeff$, decreasing correspondingly. Quantitative measures of the macroscopic curvature as a function of measured air pressure are presented in the main figure of (f); different colours indicate different experimental runs and different shapes indicate different Young's modulus, $E$, as given in the key. The predicted effective curvature from the two-dimensional model developed here is plotted as the black dashed curve.  \label{fig:3Dprint}}
\end{figure}

To demonstrate this novel route for the control of curvature, we consider experiments with a 4-cell structure based on our design with an asymmetry in the clamping angles between the top and bottom surface walls; this structure is 3D-printed in Agilus-30 and inflated using a syringe while lying horizontally on a surface (see Fig.~\ref{fig:3Dprint}). The pressure is measured using a digital manometer (RS PRO RS-8890, Radio Spares, UK). The modulus of Agilus-30 is known to be highly sensitive to temperature so experiments were performed in environments with different ambient temperatures in the range $(\SI{19}{\degreeCelsius},\SI{24}{\degreeCelsius})$; this temperature was measured using an infrared thermometer and the corresponding Young's modulus was inferred from separate experiments in a tensile tester (Yunlan Zhang, private communication). The effective curvature of the device at different pressures was inferred by measurements from images of the inflation.

Experimental measurements of the effective curvature at different air pressures are compared to the predictions of the two-dimensional model in  Fig.~\myref{fig:3Dprint}{f}. These results confirm that the effective curvature of the device decreases with increased pressure, as expected, though the magnitude of the decrease is larger than expected at the measured pressures. This agreement is rather satisfactory given the variability in modulus according to print runs and the three-dimensional nature of the experimental setup that differs from our simplified, two-dimensional model. Our analytical framework may thus provide a starting point and/or an alternative to more  intricate numerical techniques, such as finite element simulations, which are often used to model such systems.

\vspace{5mm}

\section{Discussion and conclusions}

In this paper, we have presented a simple model through which the turgor-induced mechanics of monolayer cellular sheets can be understood. We began by seeking to understand whether the familiar influence of turgor on the ability of plant leaves to resist the effects of gravity applies also to structures that are a single cell thick and do not have vasculature. We developed a simple model of a two-dimensional monolayer to determine the resistance to bending that might be expected to emerge from the turgor-induced elastic deflection of the cell walls. (This two-dimensional model is expected to provide a lower bound on the resistance to bending of such structures since curvature in the neglected third-dimension is expected to give a geometry-induced rigidity \cite{Taffetani2019}.) This allowed us to consider scenarios in which the cells have sharp corners, as is suggested by cross-sections of moss leaves (see Fig.~\myref{fig:symmetric_model}{a}), but which are difficult to describe using shell theory, for example. Our model was able to quantitatively reproduce previous observations of turgor-induced shrinkage in cell monolayers \cite{Hofhuis2016}, and shows that most moss leaves are small enough that the enhancement of their rigidity by turgor is relatively small.

Given the relative unimportance of turgor pressure for providing mechanical support against gravity in moss leaves, we also considered the curling/uncurling that is observed as water content decreases/increases. In particular, we investigated the effect of different asymmetries in the cellular structure on the curling that results from a decrease in pressure (corresponding to desiccation), focussing on two types of asymmetries: mechanical (\eg~asymmetries in the elastic moduli) and geometrical (\eg~asymmetries in wall lengths/angles). We found that mechanical asymmetries only lead to curvature in turgid conditions, with the structure remaining symmetric when flaccid. Conversely, we found that geometrical asymmetries lead to a natural curvature in flaccid conditions, but that this curvature is ironed out in turgid conditions. Since this is in keeping with observations of moss leaves (which we demonstrated in the context of \emph{Physcomitrium patens} moss leaves in Fig.~\ref{fig:MossExpts}), we conclude that the underlying cause of the leaf-curling observed biologically is likely to be a geometrical, rather than mechanical, asymmetry. However, we emphasize that our model did not account for variations of the underlying physical properties, and their anisotropy, with water content --- an effect that is well known in paper, for example \cite{Reyssat2011,Bosco2015}.

Finally, we illustrated the importance, and potential utility, of introducing a cell-level asymmetry in inflatable structures by presenting a pneumatically inflated structure whose curvature mimics that of moss leaves as the internal pressure changes: at low pressure (flaccid) the structure is curved, while at high pressure (turgid) the structure flattens out. Such a structure may have applications in soft robotics where it might represent a gripper in which internal pressurization is only required to release an object, rather than to keep hold of it. This would reduce the risk of damage induced by over-pressurization \cite{Pontin2024}, and would provide yet another example for plant-inspired soft robotics \cite{Must2019,Guo2024}. We note that extending this model system to multi-cell structures may provide additional insights into more complicated, non-vascular, plant structures or alternative mechanisms for the strain-dependent properties of certain plant tissues \cite{Lipchinsky2013}. However, such extensions of our model may require the inclusion of three-dimensional effects.

\begin{acknowledgments}
We are grateful to Yunlan Zhang for sharing her data on the modulus of Agilus30 at different ambient temperature; Yoan Coudert for providing the \emph{Physcomitrium patens}; and Madeleine Seale for imaging trials. The research leading to these results has received funding from the Leverhulme Trust (DV), the EPSRC Programme Grant `From Sensing to Collaboration' EP/V000748/1 (PM), and Agence Nationale de la Recherche, GrowFlat, ANR-21-CE30-0039-01 (AB).
\end{acknowledgments}

\appendix

\section{Independence of infinitesimal bending modulus on pressure }\label{app:Independence}

\begin{figure}[ht!]
\centering
\includegraphics[width=\linewidth]{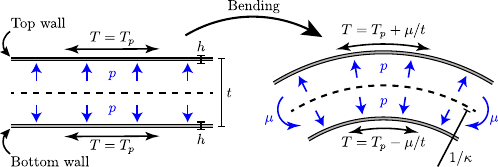}
\caption{Sketch of the bending of a pressurized two-dimensional channel. In the left sketch, an interior pressure, $p$, induces a tension, $T_p$, in the two bounding surface walls. In the right sketch, the pressurized channel is bent under a moment, $\mu$, which induces a differential strain across the channel. \label{fig:balloonparadox}}
\end{figure}

Given the everyday observation that pressure is integral to the rigidity of structures from plant leaves to cylindrical balloons, the claim that the resistance to bending at small curvatures is independent of pressure is surprising and generally involves a detailed calculation \cite{Haseganu1994,Qiu2021}. For a more intuitive feeling, consider bending a pressurized two-dimensional section through a pressurized shell, as sketched in Fig.~\ref{fig:balloonparadox}. Here, the two-dimensional section consists of two surface walls (of thickness $h$ and Young's modulus $E$) held parallel to each other with a vertical separation $t$. Pressurizing the interior induces a tension (thickness-integrated stress), $T_p$, in the two walls; for instance, for a cylindrical balloon of radius $R$, we would expect $T_p=pR/2$. Bending the section downwards to form a midline curvature, $\kappa$, induces a differential strain across the section, which stretches the top surface wall and compresses the bottom surface wall by a strain $\varepsilon= t\kappa/2$. Assuming Hooke's law, the tension in the top and bottom surface walls  then becomes $T= T_p\pm E h t\kappa/2$, respectively. Notably, the moment (about the midline) required to impose such a deformation, $\mu=Eh t^2 \kappa/2$, corresponds to $\Beff = Eh t^2/2$. This $\Beff$ is independent of $T_p$ and, hence, the internal pressure, $p$.

\section{Further size limits for monolayer leaves}\label{app:further_lim}

In \S\ref{sec:sizelimits}, we introduced the elastogravitational length, $\ellbg$, of the cellular structure. We showed that $\ellbg$ increases with the turgor pressure, $p$, in monolayer moss leaves and, for certain species, turgor is required for the leaf deflection, $\mathcal{D}\approx \ell^3/(2\ellbg)^3$, to be sufficiently small that the leaf supports itself under gravity. An alternative constraint on leaf growth might be the requirement that the differential strain caused by bending ($\ie~\mathcal{E}= \ellthick \kappa/2$, where $\kappa$ is the curvature of the leaf and $\ellthick$ is its macroscopic thickness) is small.

To estimate the curvature, $\kappa$, and, hence, the strain, $\mathcal{E}$, we use the small deflection profile of a Euler--Bernoulli beam, \ie\
\begin{equation}
w(x) =\frac{1}{24 \ellbg^3} x^2\left(6\ell^2-4 \ell x+x^2\right),
\end{equation}
where $0\leq x\leq \ell$ \cite{Howell2008}. This solution was used to estimate the deflection in \eqref{eq:mathcalDE}, \ie~$\mathcal{D}\coloneqq w(\ell)/\ell$, and also provides an estimate of the strain,
\begin{equation}\label{eq:mathcalE}
\mathcal{E}\coloneqq \frac{\ellthick}{2}w''(\ell) = \frac{\ellthick\ell^2}{4\ellbg^3}.
\end{equation}
Thus, for the strain induced by bending to be sufficiently small, we require $\ell/2$ to be small in comparison to $\ellstrain\coloneqq (\ellbg^3/\ellthick)^{1/2}= [\Beff/(\rho g \ellthick^2)]^{1/2}$.

\begin{figure}[htp!]
\centering
\includegraphics[width=\linewidth]{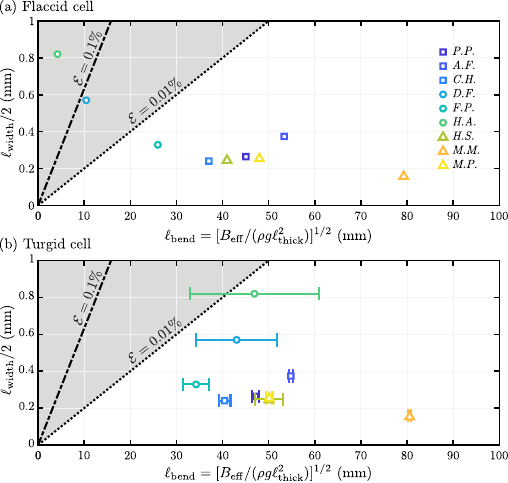}
\caption{ Plot of the extended lamina widths, $\ellwidth/2$, against estimated strain length, $\ellstrain=[\Beff/(\rho g \ellthick^2)]^{1/2}$, for nine species of single-cell thick moss, defined by colour, and point shape denoting the type: trunk-dwelling mosses, $\figwsquare$, (\textit{Pyrrhobryum pungens}, \textit{Acroporium fuscoflavum}, \textit{Campylopus hawaiicus}); ground-dwelling mosses, $\figwcircle$, (\textit{Distichophyllum freycinetii}, \textit{Fissidens pacificus}, \textit{Hookeria acutifolia}); branch-dwelling mosses, $\figwtriangle$, (\textit{Holomitrium seticalycinum}, \textit{Macromitrium microstomum}, \textit{Macromitrium piliferum}). Here, we have used the cell width, cell height, interior and surface cell wall thicknesses, lamina thickness, and maximum leaf width from Ref.~\cite{Waite2010} to estimate $L$, $\LLat$, $\hLat$, $\hSur$, $\ellthick$, and $\ellwidth$, respectively. We further assume a density $\rho=\SI{1000}{\kilogram/\metre\cubed}$; clamping angle $\Theta=\pi/4$ (an approximation based on cell photographs \cite{Waite2010}); Young's moduli $\ELat=\ESur = \SI{100}{\mega\pascal}$, which are chosen as a lower bound to the typically accepted range in plant cells \cite{Cosgrove2016}; and turgor pressures (a) $p = \SI{0}{\mega\pascal}$ (\ie~a flaccid cell) and (b) $\SI{0.8}{\mega\pascal}\leq p\leq \SI{2.1}{\mega\pascal}$ (the typical range found in moss \cite{Proctor2007}, which gives the range in $\ellstrain$). The lines $\ellwidth/2=2 \mathcal{E}^{1/2} \ellstrain$ for $\mathcal{E}= 10^{-3}$ (dash-dotted) and $\mathcal{E}= 10^{-4}$ (dotted), correspond to the maximal differential strain, $\mathcal{E}$, caused by bending a Euler--Bernoulli beam, and the shaded region corresponding to significant deformations under gravity.\label{fig:Waite2}}
\end{figure}

In Fig.~\ref{fig:Waite2}, the leaf widths, $\ellwidth$, of nine different species of single-cell thick moss leaves are plotted against the corresponding bending lengths, $\ellstrain= [\Beff/(\rho g \ellthick^2)]^{1/2}$, predicted from our model. Here, we use the same data as that used in Fig.~\ref{fig:Waite}. By comparing Fig.~\myref{fig:Waite2}{a}, for a flaccid cell, and Fig.~\myref{fig:Waite2}{b}, for a turgid cell, it is apparent that the turgor pressure increases the bending length, $\ellstrain$. In fact, using the small deflection solution, \eqref{eq:mathcalE} with $\ell= \ellwidth/2$, the expected bending-induced strains of the turgid leaves are less than $0.1\%$, \ie~$[(\ellwidth/2)/\ellstrain]^2 = 4\times\mathcal{E}<4\times10^{-3}$. However, this is not true when $p=0$ for certain species (as was the case for the elastogravitational length in Fig.~\ref{fig:Waite}).

Overall, Figs.~\ref{fig:Waite} and \ref{fig:Waite2} provide safety factors for the surface extension of moss leaves. An additional size limit of monolayer structures is that $\ell$ must be much smaller than the length scale over which the effective beam stretches due to gravity, \ie~$\ellstretch\coloneqq \Yeff/(\rho g \ellthick)$, where $\Yeff$ is the effective stretching modulus introduced in \S II.B of Ref.~\cite{SuppMat}. However, for the moss species considered here, $\ellstretch$ is orders of magnitude larger than $\ellbg$ and $\ellstrain$, thus it does not provide a principal constraint on the surface extension of moss leaves.

\section{Experiments with \emph{Physcomitrium patens}}\label{app:experiment}

\emph{Physcomitrium patens} (Gransden strain) was grown on medium (0.2\%, Gamborg’s B5 media with vitamins, Duchefa) mixed with agar (1.2\% plant agar, Duchefa), a phytotron (Aralab) with $\SI{24}{\hour}$ lighting ($\sim\SI{5300}{\lux}$) at $\SI{22}{\degreeCelsius}$. \emph{Physcomitrium} leaves were observed fresh and kept on a wet glass slide (enough water to moisturise the sample without drowning it) and then dried upon a $\SI{37}{\degreeCelsius}$ heating plate for about $\SI{14}{\minute}$. Wet and dry leaves were imaged with an Optical Coherence Tomography system (Ganymede-SP5, Thorlabs) using an objective  (LMS02-BB) with a $\SI{5}{\milli\meter}\times \SI{5}{\milli\meter}$ field of view and a pixel size of $\SI{10}{\micro\meter}\times\SI{10}{\micro\meter}\times\SI{1.95}{\micro\meter}$ 
(the smallest size is vertical). The ThorImage software (Thorlabs) was used for TIFF file extraction. TIFF stacks were processed with ImageJ for the extraction of curvature. Example cross-sections of  a dehydrated and hydrated  \emph{Physcomitrium patens} leaf is  shown in Fig.~\ref{fig:MossExpts} in the main text.

\bibliography{ArXivBib.bib}

\end{document}


\preprint{APS/123-QED}

\title{Supplemental material for  ``Mechanics of pressurized cellular sheets''}

\author{Thomas G. J. Chandler}
 \email{tgchandler@wisc.edu}
\affiliation{Mathematical Institute, University of Oxford, Woodstock Rd, Oxford, OX2 6GG, UK}
 \affiliation{Department of Mathematics, University of Wisconsin--Madison, Madison, WI 53711, USA}
 
 \author{Jordan Ferria} 
\affiliation{LadHyX, CNRS, Ecole Polytechnique, Institut Polytechnique de Paris, 91128, Palaiseau Cedex, France}

\author{Oliver Shorthose}
\affiliation{Department of Engineering Science, University of Oxford, Parks Road, Oxford, OX1 3PJ, UK}

\author{Jean-Marc Allain}
\affiliation{LMS, CNRS, Ecole Polytechnique, Institut Polytechnique de Paris, 91128, Palaiseau Cedex, France}
\affiliation{Institut Nationale de Recherche en Informatique et en Automatique, 91128, Palaiseau, France}

\author{Perla Maiolino}
\affiliation{Department of Engineering Science, University of Oxford, Parks Road, Oxford, OX1 3PJ, UK}

\author{Arezki Boudaoud} 
\affiliation{LadHyX, CNRS, Ecole Polytechnique, Institut Polytechnique de Paris, 91128, Palaiseau Cedex, France}

\author{Dominic Vella}%
 \email{dominic.vella@maths.ox.ac.uk}
\affiliation{Mathematical Institute, University of Oxford, Woodstock Rd, Oxford, OX2 6GG, UK}

\date{\today}
\maketitle

\section{Further details of the cellular model}\label{app:FurtherDetails}

In this section, we provide further details about the cellular model presented in the main paper. We focus, in particular, on deriving the governing equations and understanding the appearance of shrinkage upon first pressurization. For conciseness, we shall  introduce the asymmetric  and symmetric cell models simultaneously.

\subsection{Governing equations}\label{app:governingeqs}

In  the main text, we modelled the bending of the  cell by some applied moment, $\mu$. To model the stretching of the  cell, which is required to formulate the effective stretching modulus in  \S\ref{app:StretchingModulus}, an additional force, $\tau$, between neighbouring cells must be introduced. We assume  $\tau$  acts normal to the lateral wall, uniformly distributed along its length. Applying a force and torque balance on the cell walls yields the horizontal and vertical forces between the surface  and lateral walls,
\begin{subequations}\label{eq:NTB_TTB}
\begin{gather}
\TTB = \mp\frac{p}{2}\left(w\tan\phi\mp t \cos\phi\right)\pm\frac{\mu}{t\cos\phi}+ \frac{\tau}{2}\cos\phi
\quad \text{and} \quad \NTB = \pm \frac{p}{2}\left(w\pm t \sin\phi\right)+\frac{\tau}{2}\sin\phi,\subtag{a,b}
\end{gather}
\end{subequations}
respectively, here, and throughout, `$+$' refers to top wall and `$-$' to the bottom wall. We also distinguish between the Young's moduli and thicknesses of the top,  bottom, and lateral walls by taking $\{E,h\}\mapsto\{E_+,h_+\}$, $\{E_-,h_-\}$, and $\{\ELat,\hLat\}$,  respectively. This is to reflect observations in nature; for example, in single-cell thick moss leaves, the `interior wall thickness' and `surface wall thickness' are often quite different, with $0.3\lesssim h_L/h_\pm\lesssim 1.2$ depending on the species \cite{Waite2010}.

\subsubsection{Deformation of the lateral walls}

 Since the force  imposed by the neighbouring cells, $\tau$, is normally distributed, the lateral walls only stretch longitudinally due to a constant tension (thickness-integrated stress) $\tilde{T}\coloneqq \pm\NTB \cos\phi\mp\TTB\sin\phi$. Hence, assuming Hooke's law, $\tilde{T}=Y_L(t/L_L-1)$, the deformed length $t$ satisfies the quadratic equation derived in the main text:
\begin{equation}\label{eq:t_2}
\frac{t}{\LLat} =1\pm\frac{1}{\YLat} \left(N_\pm \cos\phi - T_\pm\sin\phi\right)= 1+\frac{1}{\YLat}\left(\frac{p w}{2\cos\phi}- \frac{\mu \tan\phi}{t} \right),
\end{equation}
where $\YLat\coloneqq\ELat\hLat$ is the stretching modulus and $L_L$ is the relaxed length of the lateral walls.

\subsubsection{Deformation of the surface walls}

Along the top (`$+$') and bottom (`$-$') surface walls, the tangential and perpendicular resultant forces, $\tilde{T}_\pm(S_\pm)$ and $\tilde{N}_\pm(S_\pm)$, and resultant moment around the $y$-axis, $\tilde{M}_\pm(S_\pm)$, satisfy the force and moment balance,
\nolinebreak
\begin{subequations}\label{eq:Kirchhoffforcemom}
\begin{gather}
\frac{1}{\alpha_\pm}\frac{\de}{\de S_\pm} \big[\tilde{T}_\pm\bm{t}_\pm-\tilde{N}_\pm \bm{n}_\pm\big]=  \pm p \bm{n}_\pm \quad\text{and}\quad
\frac{1}{\alpha_\pm}\frac{\de \tilde{M}_\pm}{\de S_\pm} = - \tilde{N}_\pm,\subtag{a,b}
\end{gather}
\end{subequations}
for $0\leq S_\pm\leq L_\pm$, where $\bm{t}_\pm=(\cos\theta_\pm,\sin\theta_\pm)$ is the unit tangent vector, $\bm{n}_\pm=(\sin\theta_\pm,-\cos\theta_\pm)$ is the unit normal vector, and $\pm p$ refers to the pressure being applied from below or above, respectively. The wall positions, $(x_\pm,z_\pm)$, are given by the geometric conditions
\begin{subequations}\label{eq:Kirchhoffgeom2}
\begin{equation}
\frac{\de x_\pm}{\de S_\pm} = \alpha_\pm \cos \theta_\pm\quad \text{and} \quad \frac{\de z_\pm}{\de S_\pm} = \alpha_\pm \sin \theta_\pm. \subtag{a,b}
\end{equation}
\end{subequations}
To close the system, we introduce two constitutive laws: Hooke's law (\ie~a linear stress--strain relationship) and a linear relationship between the change in curvature and the bending moment, that is
\begin{subequations}\label{eq:constitutive}
\begin{gather}
\tilde{T}_\pm(S_\pm)= Y_\pm \left[\alpha_\pm(S_\pm)-1\right]\quad\text{and}\quad  \tilde{M}_\pm(S_\pm) = B_\pm\left(\frac{\de\theta_\pm}{\de S_\pm}-\kappa^0_\pm\right),\subtag{a,b}
\end{gather}
\end{subequations}
respectively, where $\kappa^0_\pm$ are the intrinsic wall curvatures, and  $Y_\pm \coloneqq E_\pm h_\pm$ and $B_\pm \coloneqq E_\pm h_\pm^3/12$ are the stretching and bending moduli of the surface walls.

Integrating \eqref{eq:Kirchhoffforcemom} with \eqref{eq:Kirchhoffgeom2}, \myeqref{eq:constitutive}
{a},  and $x(0)=z(0)=0$ yields the Kirchhoff equations:
\begin{subequations}\label{eq:Kirchofffinal2}
\begin{align}
\frac{B_\pm}{\alpha_\pm}\frac{\de^2 \theta_\pm}{\de S_\pm^2} &=\TTB \sin\theta_\pm- \NTB \cos\theta_\pm \pm p\left(x_\pm\cos\theta_\pm+z_\pm\sin\theta_\pm\right),\\
\text{and} \quad Y_\pm(\alpha_\pm-1) &=\TTB \cos\theta+\NTB\sin\theta \pm p \left(z_\pm \cos \theta_\pm -x_\pm \sin \theta_\pm\right),
\end{align}
\end{subequations}
for the end forces, $(\TTB,\NTB)$,   given in \eqref{eq:NTB_TTB}. Overall, the walls' position, $(x_\pm,z_\pm)$, and intrinsic equation, $\theta_\pm(S_\pm)$, are given by the system of ODEs \eqref{eq:constitutive} and \eqref{eq:Kirchofffinal2}. This is subject to boundary conditions
\begin{subequations}\label{eq:lateralbcs2}
\begin{gather}
x_\pm(0)=0,\quad z_\pm(0)=0, \quad \theta_\pm(0) =\pm\left(\frac{\pi}{2}-\Theta_\pm\right)+\phi, \quad
x_\pm(L_\pm/2) = \frac{w}{2} \pm \frac{t}{2}\sin\phi, \quad \theta_\pm(L_\pm/2) = 0, \subtag{a--e}
\end{gather}
\end{subequations}
 with $t$ the lateral wall length given by \eqref{eq:t_2}.

\subsubsection{Dimensionless system}\label{app:nondim}

We define the dimensionless variables as
\begin{subequations}\label{eq:dimensionless}
\begin{equation}
\xi\coloneqq \frac{S_\pm}{L_\pm}, \quad X_\pm(\xi) \coloneqq \frac{x_\pm(S_\pm)}{L_\pm}, \quad Z_\pm(\xi) \coloneqq \frac{z_\pm(S_\pm)}{L_\pm},\subtag{a--c}
\end{equation}
\end{subequations}
 and consider the axial stretch, $\alpha_\pm$, and wall inclination angle, $\theta_\pm$, as functions of the dimensionless arclength, $\xi$, instead of its dimensional counterparts, $S_\pm$.

Using \eqref{eq:dimensionless}, we arrive at a dimensionless system of ODEs for the surface wall slopes, $\theta_\pm(\xi)$, and  wall positions, $(X_\pm,Z_\pm)$: 
\begin{subequations}\label{eq:ODEdimensionless}
\begin{gather}
\frac{\de X_\pm}{\de \xi} = \alpha_\pm \cos \theta_\pm, \quad
\frac{\de Z_\pm}{\de \xi} = \alpha_\pm\sin \theta_\pm,\subtag{a,b}\\
\frac{\hatb_\pm}{\alpha_\pm}\frac{\de^2 \theta_\pm}{\de \xi^2} = \hatTz \sin\theta_\pm- \hatNz \cos\theta_\pm \pm \hatp_\pm \left(X_\pm\cos\theta_\pm+Z_\pm\sin\theta_\pm\right), \subtag{c}
\end{gather}
\end{subequations}
for $0\leq \xi\leq 1$, with the axial stretch
\begin{equation}\label{eq:ODEaxial}
\alpha_\pm =1+ \hatTz \cos\theta+\hatNz\sin\theta \pm \hatp_\pm  \left(Z_\pm \cos \theta -X_\pm \sin \theta\right),
\end{equation}
the end forces
\begin{subequations}\label{eq:hatTs}
\begin{align}
\hatTz \coloneqq \frac{\TTB}{Y_\pm}& =\frac{\hatp_\pm}{2}\left( \A_\pm\hatt \cos\phi\mp\hatw_\pm\tan\phi\right)\pm\frac{\hatmu_\pm}{\hatt\cos\phi}+\frac{\hattau_\pm}{2}\cos\phi,\\
\text{and}\quad
\hatNz \coloneqq \frac{\NTB}{Y_\pm} &= \frac{\hatp_\pm}{2}\left( \A_\pm\hatt \sin\phi \pm \hatw_\pm\right)+\frac{\hattau_\pm}{2}\sin\phi,
\end{align}
\end{subequations}
and the  lateral wall stretch $\hatt\coloneqq t/\LLat$, which solves
\begin{equation}\label{eq:hatt}
\hatt = 1+\frac{1}{\hatY_\pm}\left(\frac{\hatp_\pm \hatw_\pm}{2\cos\phi}- \frac{\hatmu \tan\phi}{\hatt} \right).
\end{equation}

Solutions to \eqref{eq:ODEdimensionless} are subject to
\begin{subequations}\label{eq:BCdimensionless}
\begin{gather}
 X_\pm(0)=0,\quad Z_\pm(0)=0, \quad \theta_\pm(0) =\pm\left(\frac{\pi}{2}-\Theta_\pm\right)+\phi,\quad X_\pm(1/2) =\frac{\hatw_\pm}{2} \pm\frac{\A_\pm}{2}\hatt\sin\phi,\quad \theta_\pm(1/2) =0.\subtag{a--e}
\end{gather}
\end{subequations}
Here, the extra boundary conditions for $X_\pm$ determines the unknown lateral wall angle, $\phi$, and cell width, $\hatw_\pm\coloneqq w/L_\pm$. 

The system \eqref{eq:ODEdimensionless}--\eqref{eq:BCdimensionless} is governed by the dimensionless parameters
\begin{subequations}\label{eq:dimensionlesspars}
\begin{gather}
\hatp_\pm \coloneqq \frac{p L_\pm}{Y_\pm}, \quad
\A_\pm \coloneqq \frac{\LLat}{L_\pm}, \quad
\hatb_\pm\coloneqq \frac{B_\pm}{Y_\pm L_\pm^2},\quad
\hatY_\pm \coloneqq \frac{\YLat}{Y_\pm}, \quad
\hatmu_\pm \coloneqq \frac{\mu}{Y_\pm\LLat},\quad
\hattau_\pm \coloneqq \frac{\tau}{Y_\pm}. \subtag{a--f}
\end{gather}
\end{subequations}
Note that $\hatp_+=(\mathcal{R}_E\mathcal{R}_h/\mathcal{R}_L)\hatp_-$,  $\A_+=\mathcal{R}_L\A_-$, $\hatb_+=(\mathcal{R}_L/\mathcal{R}_h)^2\hatb_-$, $\hatY_+=(\mathcal{R}_E\mathcal{R}_h)\hatY_-$, $\hatmu_+ =(\mathcal{R}_E\mathcal{R}_h) \hatmu_-$, and $\hattau_+ =(\mathcal{R}_E\mathcal{R}_h) \hattau_-$ for the asymmetric parameter ratios
\begin{subequations}
\begin{gather}
\mathcal{R}_E\coloneqq \frac{E_-}{E_+}, \quad \mathcal{R}_h\coloneqq \frac{h_-}{h_+}, \quad \mathcal{R}_L\coloneqq \frac{L_-}{L_+},\quad  \mathcal{R}_\Theta\coloneqq \frac{\Theta_-}{\Theta_+}.\subtag{a--d}
\end{gather}
\end{subequations}
The symmetric cell model considered in \S II in the main text takes  $\tau= 0$ (no applied tension), $\hatY_\pm=1$ (a homogeneous cell), and $\mathcal{R}_E=\mathcal{R}_h=\mathcal{R}_L=\mathcal{R}_\Theta=1$ (an up--down symmetric cell); hence, the  `$\pm$'  subscripts in the dimensionless parameters \eqref{eq:dimensionlesspars} can be dropped. The asymmetric cell model  in \S III of the main text takes $\tau=\mu= 0$ (no applied forcing).

\subsection{The symmetric base case: no external loads\label{sec:BaseCase}}

 We first consider the behaviour of the symmetric system under no applied external load, $\tau=\mu=0$. As symmetric, we drop the `$\pm$' in the dimensionless variables and use subscript `$0$' to denote no applied load.

\paragraph{Flaccid cell $(\hatp\ll\hatb)$} For a flaccid cell, $\hatp\ll\hatb$ and so we take $\hatp \approx0$ with no end loads, $\hattau=\hatmu=0$. The system then simplifies considerably: $\hatTz=\hatNz=0$  from \eqref{eq:hatTs}, and so \myeqref{eq:ODEdimensionless}{c} may be integrated subject to  \myeqref{eq:BCdimensionless}{c,e}. This yields the intrinsic equations of the surface walls,
\begin{equation}\label{eq:flaccidTheta0}
\theta_\pm(\xi)= \pm \left(\frac{\pi}{2}-\Theta\right)(1-2\xi).
\end{equation} 
The corresponding walls are in the shape of circular segments, which  may be found formally by integrating \myeqref{eq:ODEdimensionless}{a,b} subject to \myeqref{eq:BCdimensionless}{a,b}. The cell  then consists of two circular segments  attached to a rectangular centre of thickness, width, and area
\begin{subequations}\label{eq:flaccidtwA}
\begin{gather}
\hatt_0=1,\quad \hatw_0 = \frac{2\cos \Theta}{\pi-2\Theta},\quad
\hatA_0 = \frac{2\cos\Theta}{\pi-2\Theta} +  \frac{\pi-2\Theta-\sin\left(2\Theta\right)}{\A\left(\pi-2\Theta\right)^2}, \subtag{a--c}
\end{gather}
\end{subequations}
respectively. Here the cell width increases monotonically with the clamping angle, $\Theta$, whilst the cell area is maximized at $\Theta=\Thetam$, which is defined in Eq.~(7) of the main text. We note that $\Thetam$ can be approximated as $\Thetam\approx \pi\A/[2(1+\A)]$, with a relative error of less than $7\%$ for all $\A$.

\paragraph{Turgid cell $( \hatb\ll \hatp\ll 1)$} For a turgid cell, $ \hatb\ll\hatp$ (but with $\hatp\ll1$ to ensure small deformations), the right-hand side of \myeqref{eq:ODEdimensionless}{c} with $\hatmu=\hattau=0$, must vanish; hence, we find that 
\begin{equation}
\left(X_\pm(\xi)-\frac{\hatw_0}{2}\right)^2+ \left(Z_\pm(\xi)\pm\frac{\A\hatt_0}{2}\right)^2 = \hatR_0^2 \coloneqq \frac{\hatw_0^2+(\A\hatt_0)^2}{4},
\end{equation}
and $\alpha_\pm(\xi)=1+\hatp\hatR_0$. Here, the clamping conditions, \myeqref{eq:BCdimensionless}{c}, are only apparent in boundary layers of width ${\xi=\Oh(\hatb/\hatp)}$, and so the cell resembles a circle (of radius $R_0= L\hatR_0$) with two removed segments (of chord length $t_0= \LLat\hatt_0$) corresponding to the two lateral walls. From elementary geometry, the total deformed length of the surface walls, \ie~$\int_0^1\alpha_\pm\de\xi$, must equal $\hatR_0(\pi- \gamma_0)$, where $\gamma_0\coloneqq2\arctan(\hatt_0\A/\hatw_0)$ is the angle of the removed segments. Consequently, the unknown cell width, $\hatw_0$, and thickness, $\hatt_0$, are found by solving
\begin{subequations}\label{eq:R0t0w0}
\begin{gather}
\int_0^1\alpha_\pm(\xi)\de\xi= 1+\hatp\hatR_0= \hatR_0\left(\pi-\gamma_0\right)  \quad \text{and} \quad \hatt_0=1+\hatp\frac{\hatw_0}{2\hatY}. \subtag{a,b}
\end{gather}
\end{subequations}
The cell area is given by
\begin{equation}
\hatA_0 =\frac{\hatw_0\hatt_0}{2}+\frac{\hatR_0^2}{\A}\left(\pi-\gamma_0\right).
\end{equation}

\paragraph{Moderate internal pressures} At moderate internal pressures, the cell thickness $\hatt_0$, width $\hatw_0$, and area $\hatA_0$, must be determined numerically. Figure~2 of the main text, shows the numerically determined $\hatw_0$ and $\hatA_0$ as functions of the internal pressure, $\hatp$, for bending stiffness $\hatb=10^{-4}$, aspect ratio $\A=1/2$, $\hatY=1$, and various clamping angles, $\Theta$. Note that $\hatt_0$ increases (approximately) linearly with $\hatp$  and is not plotted. In the inset of this figure, we observe the non-monotonic dependence of the flaccid cell area on the clamping angle, $\Theta$, \ie~\myeqref{eq:flaccidtwA}{c}: $\hatA_0$ is maximized at $\hatp=0$ when $\Theta=\Thetam\approx 0.54$, for which $\hatA_0\approx 1.39$. Thus, $\Theta>\Thetam$ for $\Theta=\pi/2$, $3\pi/8$, and $\pi/4$,  so the cell width, $\hatw_0$, initially decreases with $\hatp$; while $\Theta<\Thetam$ for $\Theta=\pi/8$ and $0$, so  $\hatw_0$ initially increases with $\hatp$.

Varying the other cell parameters has the following effects, $\A$ alters the critical angle, $\Thetam$; $\hatb$ changes the onset of the turgid regime ($ \hatp\gg\hatb$); and $\hatY$ controls the size of the lateral walls, which only quantitatively changes the results presented here.

\section{Effective tissue parameters \label{app:EffectivePars}}

In this section, we provide further details of the effective macroscale properties derived from our cellular model. For the symmetric cell, we consider the effective bulk modulus ($\Keff$) in \S\ref{app:BulkModulus}, the effective stretching modulus ($\Yeff$) in \S\ref{app:StretchingModulus}, and the effective bending modulus ($\Beff$) in \S\ref{app:BendingModulus}. For the asymmetric cell, we compute the effective curvature ($\kapeff$) in \S\ref{app:Curvature}. We shall  be able to determine asymptotic results valid for $\hatp\ll\hatb$ by setting $\hatp\approx0$. We refer to such cells as \emph{flaccid}. Similarly, asymptotic progress is often possible when $\hatb\ll\hatp$ (but ensuring that $\hatp\ll1$ to avoid the effects of nonlinear elasticity). We refer to such cells as \emph{turgid}.

\subsection{Effective bulk modulus}\label{app:BulkModulus}

As discussed in \S II.B of the main text, the cell area of the symmetric cell, 
\begin{equation}\label{eq:area}
\hatA_0\coloneqq \frac{A_0}{\LLat L}
= \hatw_0\hatt_0+ \frac{1}{\A}\int_0^1\left( \ZT\frac{\de \XT}{\de\xi}-\ZB\frac{\de \XB}{\de\xi}\right)\de\xi,
\end{equation} 
increases as the internal pressure increases: turgor pressure drives cell expansion. This increase in area is caused by a change in cell thickness $\hatt_0$, cell width $\hatw_0$, and the deformation of the surface walls. For an isotropic elastic body, the change in volume with pressure is commonly measured through a bulk modulus, $K=-V \de P/\de V$, where $P$ is the imposed external pressure and $V$ is the volume of the substance. An analogous \emph{effective 2D bulk modulus} can be introduced for the two-dimensional cell as
\begin{equation}\label{eq:Keff}
\Keff = A_0 \frac{\de p}{\de A_0}.
\end{equation}
Note  the sign change between $K$ and $\Keff$ occurs because of the switch from an external to internal pressure.

To non-dimensionalize the effective bulk modulus, $\Keff$, we use the unloaded, flaccid ($\hatp\ll\hatb$), cell limit studied in \S\ref{sec:BaseCase}, with $\Theta=\pi/2$ as the reference state. For $\Theta=\pi/2$, the wall angle \eqref{eq:flaccidTheta0} vanishes as $\hatp\to0$; it follows that the corresponding enclosed area is
\begin{equation}\label{eq:areaexpandpi2}
\hatA_0 = 1 + \left(\frac{\A}{2}+\frac{1}{2\hatY}+ \frac{1}{360\A\hatb}\right)\hatp+\Oh(\hatp^2),
\end{equation}
as $\hatp\to0$. At leading-order, $\hatA_0= 1$, due to the choice of dimensionless area. The three parts of the $\Oh(\hatp)$-term describe the initial increase in area caused by the stretching of the surface walls, stretching of the lateral walls, and bending of the surface walls (which is most prominent since $\hatb\ll1$), respectively. Inserting \eqref{eq:areaexpandpi2} into \eqref{eq:Keff} gives the bulk modulus
\begin{equation}
\Keff^0 \coloneqq \frac{Y}{L} \times \left(\frac{\A}{2}+\frac{1}{2\hatY}+ \frac{1}{360 \hatb\A}\right)^{-1},
\end{equation}
for $\hatp=0$ and $\Theta=\pi/2$. We, thus, define the dimensionless effective bulk modulus as $\hatKeff\coloneqq \Keff/\Keff^0$.

\begin{figure}[ht!]
\centering
\includegraphics[width=.75\linewidth]{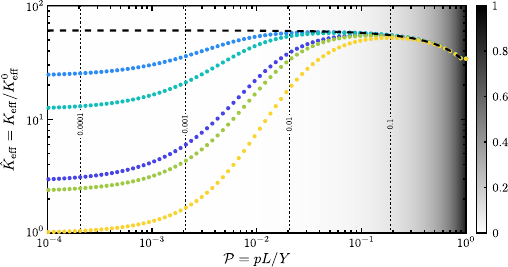}
\caption{Plot of the numerically-determined bulk modulus, $\hatKeff$, as a function of the internal pressure, $\hatp$, for the configuration parameters $\hatY=1$, $\A=1/2$, $\hatb=10^{-4}$, and $\Theta=\pi/2$, $3\pi/8$, $\pi/4$, $\pi/8$, and $0$ (yellow to blue dots). Plotted as a black dashed curve is the circular cell solution with $\Theta=\Thetam\approx 0.54$. The greyscale background corresponds to the maximum strain in the top and bottom beams, $\epsilon_{\max}\coloneqq\max_{\xi}|\alpha_\pm-1|$, as delineated by the colourbar; the contour lines of $\epsilon_{\max}=0.0001$, $0.001$, $0.01$, and $0.1$ are shown as vertical dashed lines. For small wall strains, the increase in pressure is resisted more as the cell tends towards the circular configuration (which is achieved when $\Theta=\Thetam$, for all $\hatp$).\label{fig:Keff} }
\end{figure}

The $\hatKeff$ determined numerically are plotted as a function of $\hatp$ in Fig.~\ref{fig:Keff}. This plot shows that, for a fixed $\Theta\neq \Thetam$ and $\hatp\ll1$, there is an increasing resistance to an  increase of pressure (\ie~$\hatKeff$ increases with $\hatp$). In the critical case, $\Theta=\Thetam$ with $\hatp\ll 1$, the effective bulk modulus acts significantly differently to the other cases: the cell is circular for all internal pressures, so the enclosed area is already maximized and $\hatKeff$ is approximately constant.  For $\hatp\sim 1$, the pressure is large enough to stretch the cell walls, and so, $\hatKeff$ decreases.

\subsection{Effective stretching modulus} \label{app:StretchingModulus}

With an additional applied tension, $\hattau$, but maintaining no applied moment, $\hatmu=0$, the pressurized symmetric cell is expected to stretch longitudinally, leading to some strain $\epsilon$. The cell width takes the form $\hatw = \hatw_0(1+\epsilon)$, where $\hatw_0$ is the width of the cell with no additional strain, $\hattau=0$. The up--down symmetry is not broken by pure stretching and the angle of the lateral walls must vanish, $\phi= 0$. The resulting system yields a linear relationship between the applied tension and the midline strain at $\Oh(\epsilon)$ as $\hattau\sim\epsilon\to 0 $, and so, we introduce an \emph{effective stretching modulus} 
\begin{equation}
\Yeff \coloneqq \lim_{\tau\to0}\tau/\epsilon.
\end{equation}

To non-dimensionalize $\Yeff$, we use the stiffness of a flaccid cell ($\hatp=0$) with $\Theta=\pi/2$. The top and bottom surface walls are stretched longitudinally when  $\Theta=\pi/2$ and $\hatp=0$, resulting in an  effective stretching modulus  $\Yeff^0 \coloneqq 2 Y$. The dimensionless effective stretching modulus is  defined as $\hatYeff\coloneqq \Yeff/\Yeff^0$. In general, $\hatYeff$ must be determined numerically (we use \texttt{bvp4c} in \texttt{\textsc{Matlab}}), however analytical progress can be made in the asymptotic limits of large and small pressures, which we consider now.

\paragraph{Flaccid cell} Solving \eqref{eq:ODEdimensionless}--\eqref{eq:BCdimensionless} with $\hatmu=\phi=\hatp=0$ using an asymptotic expansion as $\hattau\sim\epsilon\to0$, we find the intrinsic equation of the surface walls
\begin{equation}
\theta_\pm(\xi)\sim \theta_0(\xi) \pm\frac{\hattau}{2\hatb(\pi-2\Theta)^2}\Big\{(1-2\xi)\cos\Theta- \sin\left[\left(\pi/2-\Theta\right)(1-2\xi)\right]\Big\},
\end{equation}
and the axial stretch
\begin{equation}
\alpha_\pm(\xi) \sim 1 + \frac{\hattau}{2} \cos\left[\left(\pi/2-\Theta\right)(1-2\xi)\right],
\end{equation}
where $\theta_0(\xi)$ is the leading-order solution in \eqref{eq:flaccidTheta0}. Inserting these expressions into \myeqref{eq:ODEdimensionless}{b} and integrating over the domain, $0\leq \xi \leq 1/2$, yields an expression for the stretched cell length, $2X_\pm(1/2)=\hatw_0(1+\epsilon)$. At $\Oh(\epsilon)$, this results in a linear tension--strain relationship,  yielding the effective stretching modulus 
\begin{equation}\label{eq:smallPYeff}
\hatYeff\sim \frac{4\hatb\cos\Theta(\pi-2\Theta)^3}{\left[1+\hatb(\pi-2\Theta)^2\right](\pi-2\Theta)(\pi-2\Theta+\sin2\Theta)-4(1+\cos2\Theta)}.
\end{equation} 

\begin{figure}[ht!]
\centering
\includegraphics[width=.75\linewidth]{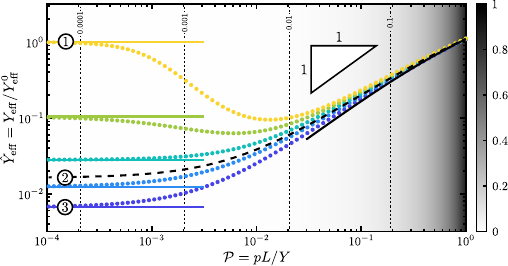}
\caption{Plot of the effective stretching modulus, $\hatYeff$, as a function of the internal  $\hatY=1$, $\A=1/2$, $\hatb=10^{-4}$, and $\Theta=\pi/2$, $3\pi/8$, $\pi/4$, $\pi/8$, and $0$ (yellow to blue). The $\hatYeff$ found by numerically solving the Kirchhoff equations, \eqref{eq:ODEdimensionless}--\eqref{eq:BCdimensionless} with $\hatmu=0$ and $\hattau\sim\epsilon\ll1$, are plotted as coloured dots, whilst asymptotic results are plotted as solid curves: \eqref{eq:smallPYeff}, valid for small pressures, is shown in colour and \eqref{eq:largePYeff}, valid for large pressures, is shown in black. Plotted as a black dashed curve is the numerical solution with maximum area, $\Theta=\Thetam\approx 0.54$.  \ding{172}--\ding{174} correspond to three regimes sketched in Figure 2 of the main text.  The greyscale background corresponds to the maximum strain in the top and bottom beams, $\epsilon_{\max}\coloneqq\max_{\xi}|\alpha_\pm-1|$, as delineated by the colourbar; the contour lines of $\epsilon_{\max}=0.0001$, $0.001$, $0.01$, and $0.1$ are shown as vertical dashed lines. Although $\hatYeff$ may intially decrease with $\hatp$ (\ie~for large values of $\Theta$), it ultimately increases with $\hatp$ according to \eqref{eq:largePYeff}. \label{fig:stretmod}}
\end{figure}

\paragraph{Turgid cell} For turgid cells, the right hand side of \myeqref{eq:ODEdimensionless}{c} must vanish. The surface walls are circular arcs with radius, $R= L\hatR$, and sector angle, $\pi-\gamma$, given by
\begin{subequations}\label{eq:largePRgamma}
\begin{gather}
4\hatR^2 =\hatw^2 + \left(\A\hatt+\hattau/\hatp\right)^2\quad
\text{and}\quad
\gamma= 2\arctan\left(\frac{\A\hatt+\hattau/\hatp}{\hatw}\right),\subtag{a,b}
\end{gather}
\end{subequations}
where $\hatw=\hatw_0(1+\epsilon)$ and $\hatt = 1+ \hatp \hatw/(2\hatY)$. The stretched turgid cell is, thus, in the shape of a symmetric lens with truncated tips. The clamped boundary conditions, \myeqref{eq:BCdimensionless}{c}, are only apparent in boundary layers of width $\Oh(\hatb/\hatp)$, we assume that these boundary layers are negligible when calculating $\hatYeff$ at large pressures. Inserting \eqref{eq:largePRgamma} into the arclength requirement, \myeqref{eq:R0t0w0}{a}, and expanding in powers of $\epsilon\sim\hattau\to 0$, yields a linear tension--strain relationship at $\Oh(\epsilon)$, giving the effective stretching modulus
\begin{equation}\label{eq:largePYeff}
\hatYeff\sim \frac{\hatp\hatw_0}{2}\left(\frac{\hatw_0+2\hatR_0\A\hatt_0}{2\hatR_0\hatw_0-\A\hatt_0} - \frac{\A\hatp}{2\hatY}\right),
\end{equation} 
where $\hatR_0^2 \coloneqq (\hatw_0^2 +\A^2\hatt_0^2)/4$, $\hatw_0$, and $\hatt_0$ are the  radius, width, and thickness of the `unstretched' cell,  which solve \eqref{eq:R0t0w0}. For small axial strains, $\hatp\ll 1$, \eqref{eq:largePYeff} increases linearly with pressure, while for large axial strains, $\hatp\sim 1$, the pressure has a sublinear effect (see Fig.~\ref{fig:stretmod}). An effective stretching modulus that is linear in the pressure has previously been found for pressurized cellular structures \cite{Gao2023}. The analytical result found by \citet{Gao2023} is recovered in the inextensible limit ($\mathcal{Y}\to\infty$) of \eqref{eq:largePYeff}, which also has the affect of removing the sublinear behaviour.

\paragraph{Moderate internal pressures} At moderate internal pressures (\ie~$\hatp\sim\hatb$) the effective stretching modulus, $\hatYeff$, must be determined numerically. In Fig.~\ref{fig:stretmod},  numerically-determined $\hatYeff$ are plotted as a function of $\hatp$. Also plotted are the asymptotic results at small and large pressures, \eqref{eq:smallPYeff} and \eqref{eq:largePYeff}, respectively. As well as showing that the expected asymptotic behaviours are recovered, Fig.~\ref{fig:stretmod} shows that $\hatYeff$ may initially decrease with pressure if the exterior clamping angle, $\Theta$, is sufficiently large, but ultimately increases with $\hatp$ according to \eqref{eq:largePYeff}. Altering the cell parameters, $\hatY$, $\A$, and $\hatb$, appear to only quantitatively affect the results presented here.

\subsection{Effective bending modulus}\label{app:BendingModulus}

In \S II.C of the main text, we defined the effective bending modulus $\Beff$ of the symmetric cell. In general, \eqref{eq:ODEdimensionless}--\eqref{eq:BCdimensionless} with $\hattau=0$ must be solved numerically  to determine $\hatBeff$ (we use \texttt{bvp4c} in \texttt{\textsc{Matlab}}). Analytical progress can be made, however, for flaccid and turgid cells; here, we provide further details about these asymptotic results.

\paragraph{Flaccid cell} Solving \eqref{eq:ODEdimensionless}--\eqref{eq:BCdimensionless} with $\hattau=0$ and $\hatw=\hatw_0+\Oh(\phi^2)$ using an asymptotic expansion as $\hatmu\sim\phi\to0$, we find the intrinsic equation of the surface walls,
\begin{subequations}
\begin{equation}
\theta_\pm(\xi)\sim \theta_0(\xi) +(1-2\zeta)\phi +\Big\{(1-2\xi)\cos\Theta - \sin\left[\left(\frac{\pi}{2}-\Theta\right)(1-2\xi)\right]\Big\}\frac{\mu}{\hatb(\pi-2\Theta)^2}\,
\end{equation}
and the axial stretch,
\begin{equation}
\alpha_\pm(\xi) \sim 1 \pm \mu\cos\left[\left(\frac{\pi}{2}-\Theta\right)(1-2\xi)\right],
\end{equation}
\end{subequations}
 where $\theta_0(\xi)$ is the leading-order solution from \eqref{eq:flaccidTheta0}. Inserting these expressions into \myeqref{eq:ODEdimensionless}{a} and integrating yields an expression for the stretched cell length, $2X_\pm(1/2)=\hatw_0\pm \A\phi/2$. This results in a linear moment--curvature relationship at $\Oh(\phi)$, which gives the effective bending modulus
\begin{equation}\label{eq:Beff_smallp}
\hatBeff \sim \hatBeff^0\coloneqq4 \hatb\frac{\cos\Theta(\pi-2\Theta)^3 +(\pi-2\Theta)\left[2(1+\cos2\Theta)- (\pi-2\Theta)\sin2\Theta\right]/\A}{(\pi-2\Theta)(\pi-2\Theta+\sin2\Theta)\left[1+\hatb(\pi-2\Theta)^2\right]-4(1+\cos2\Theta)}.
\end{equation}

\paragraph{Turgid cell} As described in \S\ref{sec:BaseCase}, for $\hatp\gg \hatb$ the right hand side of \myeqref{eq:ODEdimensionless}{c} must vanish. We, hence, find that the upper and lower surface walls are circular arcs with radius $R= L\hatR$ and sector angle $\pi-\gamma$ such that
\begin{subequations}\label{eq:largePRGamma_Beff}
\begin{equation}
\hatR^2 \sim\hatR_0^2\pm \frac{\A}{\hatp}\hatmu\quad
\text{and} \quad \gamma \sim \gamma_0 \mp 2\phi \pm \frac{\hatw_0/\hatR_0^2}{\hatt_0\hatp}\hatmu, \subtag{a,b}
\end{equation}
\end{subequations}
as $\hatmu\sim\phi\to 0 $, where $\hatR_0^2 \coloneqq (\hatw_0^2 +\A^2\hatt_0^2)/4$, $\gamma_0 \coloneqq 2\arctan(\A\hatt_0/\hatw_0)$, $\hatw_0$, and $\hatt_0$ are the radius, sector angle, width, and thickness of the `unbent' cell,  which solve \eqref{eq:R0t0w0}. 

Here, the clamped boundary conditions, \myeqref{eq:BCdimensionless}{c}, are assumed to be negligible when calculating $\hatBeff$ at large pressures since they are only apparent in boundary layers of width $\Oh(\hatb/\hatp)$. Inserting \eqref{eq:largePRGamma_Beff} into the arclength requirement, \myeqref{eq:R0t0w0}{a}, yields a linear moment--curvature relationship at $\Oh(\phi)$, which gives the effective bending modulus 
\begin{equation}
\hatBeff \sim \frac{4 \hatR_0^3 \hatt_0\hatw_0/\A}{2\hatR_0\hatw_0-\A\hatt_0}\hatp.
\end{equation}

\subsection{Effective curvature} \label{app:Curvature}

In \S III.B of the main text, we formulated the effective curvature of the cell, $\hatkapeff$, by assuming the top and bottom surface walls differ in their relaxed length $\LTB$, thickness $\hTB$, Young's modulus $\ETB$, and/or clamping angle $\ThetaTB$. In general, the dimensionless system \eqref{eq:ODEdimensionless}--\eqref{eq:BCdimensionless} with $\hatmu_\pm=\hattau_\pm=0$ must be solved numerically to determine  $\hatkapeff$ (we use \texttt{bvp4c} in \texttt{\textsc{Matlab}}). However, as was true for the symmetric cell, progress can be made in the limits of flaccid and turgid cells; we consider these limiting cases now. 

\paragraph{Flaccid cell} For a flaccid cell, integrating \myeqref{eq:ODEdimensionless}{c} with $\hatTz=\hatNz=\hatp=0$ and applying \myeqref{eq:BCdimensionless}{c,e} results in the intrinsic equations of the surface walls,
\begin{subequations}
\begin{equation}
\theta_\pm(\xi) = \pm\left(\frac{\pi}{2}-\ThetaTB\pm \phi\right)\left(1-2\xi\right).
\end{equation}
\end{subequations}
A similar result was found for the symmetric cell, \eqref{eq:flaccidTheta0}, however here the tilt of the flat wall-segments, $\phi$, does not necessarily vanish and instead needs to be determined. The corresponding wall positions may be found by integrating \myeqref{eq:ODEdimensionless}{a,b} and imposing \myeqref{eq:BCdimensionless}{a,b}. The resulting cell has width
\begin{equation}
\hatwT = \frac{\cos\left(\phi-\ThetaT\right)}{2\phi-2\ThetaT+\pi}-\RL\frac{\cos\left(\phi+\ThetaB\right)}{2\phi + 2\ThetaB-\pi},
\end{equation}
where the lateral wall angle, $\phi$, satisfies
\begin{equation}\label{eq:flaccidphi}
\AT \sin\phi= \frac{\cos\left(\phi-\ThetaT\right)}{2\phi-2\ThetaT+\pi}+\RL\frac{\cos\left(\phi+\ThetaB\right)}{2\phi + 2\ThetaB-\pi},
\end{equation}
which is solved numerically. The solutions to these equations can then be used to calculate effective curvature, $\hatkapeff= 2\AT\sin\phi/\hatwT$. Note that, here, $\hatkapeff$ is independent of $\RE$ and $\Rh$: an asymmetry in the Young's modulus or thickness of the surface walls does \emph{not} induce a curvature in the flaccid cell, which is to be expected because the surface walls are not strained for $\hatp=0$ and are only restricted by the geometry of the cell. 

\paragraph{Turgid cell} As was the case for the symmetric cell, for large internal pressures ($\hatpTB\gg \hatbTB$) the right hand side of \myeqref{eq:ODEdimensionless}{c} must vanish, and the top (`$+$') and bottom (`$-$') surface walls form circular arcs of radius $R = \LTB \hat{R}_\pm$ and sector angle $\pi-\gamma_\pm$ given by
\begin{subequations}
\begin{gather}
4\hat{R}_\pm^2 = \hatwTB^2\sec^2\phi+\hatt^2\ATB^2\quad \text{and} \quad \gamma_\pm =2\arctan\left(\frac{\ATB\hatt\cos\phi\mp\hatwTB\tan\phi}{\hatwTB\pm\ATB\hatt\sin\phi}\right).\subtag{a,b}
\end{gather}
\end{subequations}
In dimensional variables, the cell is a circle of radius $R \coloneqq (w^2\sec^2\phi+t^2)^{1/2}/2$ with two removed segments (of chord length $t$) that are oriented to form an angle $\phi$ with the vertical. Here, $\phi$, $R$, and $t$ solve equivalent conditions to \eqref{eq:R0t0w0},
\begin{subequations}\label{eq:largePasymarc}
\begin{gather}
1+\hatpTB\hat{R}_\pm = \hat{R}_\pm\left(\pi-\gamma_\pm\right) \quad
 \text{and} \quad \hatt =1+ \hatpTB\hatwTB/(2\cos\phi\hatYTB).\subtag{a,b}
\end{gather}
\end{subequations}
The cell curvature can then be computed using $\hatkapeff= 2\ATB\sin\phi/\hatwTB$.  The macroscopic curvature of a turgid cell is predominately set by the relaxed length ratio, $\RL$. However, at large axial strains ($\hatpTB\sim 1$) the pressure in \eqref{eq:largePasymarc} is non-negligible, hence $\hatkapeff$ also depends on the ratio $\RE\Rh\equiv \EB\hB/(\ET\hT)$. Note that an asymmetry in the clamping angles ($\RT\neq 1$) does \emph{not} induce a curvature in the turgid cells since the clamping conditions, \myeqref{eq:BCdimensionless}{c}, are only apparent in narrow boundary layers.

\bibliography{ArXivBib}